\documentclass[superscriptaddress,nofootinbib,twocolumn]{revtex4-1}
\usepackage{enumitem}
\usepackage{comment}
\usepackage{slashed}
\usepackage{latexsym}
\usepackage{graphics}
\usepackage{bm}
\usepackage{color}
\usepackage{amsmath}
\usepackage{amssymb}
\usepackage{graphicx}
\usepackage{subcaption}
\usepackage{slashed}
\usepackage[dvipsnames]{xcolor}

\usepackage{ezedits}
   \definecolor{r}{RGB}{255,0,0}
   \definecolor{R}{RGB}{255,0,0}
   \definecolor{o}{RGB}{255,94,1}
   \definecolor{O}{RGB}{255,165,0}
   \definecolor{p}{RGB}{223,0,225}
   \definecolor{P}{RGB}{223,0,225}
   \definecolor{g}{RGB}{0,150,85}
   \definecolor{G}{RGB}{0,150,85}
   \definecolor{b}{RGB}{0,0,255}
   \definecolor{B}{RGB}{0,0,255}
\defineEditQuick{HR}{r}
\defineEditQuick{PG}{p}

\newcommand{\eV}{\text{eV}}

\usepackage{comment}

\begin{document}

\title{Millicharged dark matter detection with ion traps}

\author{Dmitry Budker}
 \affiliation{Johannes Gutenberg-Universit{\"a}t Mainz, 55128 Mainz, Germany}
 \affiliation{Helmholtz-Institut, GSI Helmholtzzentrum f{\"u}r Schwerionenforschung, 55128 Mainz, Germany}
\affiliation{Department of Physics, University of California, Berkeley, California 94720, USA}

\author{Peter W.~Graham}
\affiliation{Stanford Institute for Theoretical Physics,
Stanford University, Stanford, CA 94305, USA}
\affiliation{Kavli Institute for Particle Astrophysics \& Cosmology, Stanford University, Stanford, CA 94305, USA}
\author{Harikrishnan Ramani}\email{hramani@stanford.edu}
\affiliation{Stanford Institute for Theoretical Physics,
Stanford University, Stanford, CA 94305, USA}

\author{Ferdinand Schmidt-Kaler}
 \affiliation{Johannes Gutenberg-Universit{\"a}t Mainz, 55128 Mainz, Germany}
 \affiliation{Helmholtz-Institut, GSI Helmholtzzentrum f{\"u}r Schwerionenforschung, 55128 Mainz, Germany}
\author{Christian Smorra}
\affiliation{Institute of Physics, Johannes Gutenberg University Mainz, Mainz, Germany}
\author{Stefan Ulmer}
\affiliation{RIKEN, Fundamental Symmetries Laboratory, Wako, Saitama, Japan}

\begin{abstract}
We propose the use of trapped ions for detection of millicharged dark matter.
Millicharged particles will scatter off the ions, giving a signal either in individual events or in the overall heating rate of the ions.
Ion traps have several properties which make them ideal detectors for such a signal.  First, ion traps have demonstrated significant isolation of the ions from the environment, greatly reducing the background heating and event rates.
Second, ion traps can have low thresholds for detection of energy deposition, down to $\sim \text{neV}$.
Third, since the ions are charged, they naturally have large cross sections for scattering with the millicharged particles, further enhanced by the low velocities of the thermalized millicharges.
Despite ion-trap setups being optimized for other goals, we find that existing measurements put new constraints on millicharged dark matter which are many orders of magnitude beyond previous bounds. For example, for a millicharge dark matter mass $m_Q=10~\textrm{GeV}$ and charge $10^{-3}$ of the electron charge, ion traps limit the local density to be $n_Q \lesssim 1 \, \textrm{cm}^{-3}$, a factor $\sim 10^8$ better than current constraints.
Future dedicated ion trap experiments could reach even further into unexplored parameter space.
\end{abstract}
\date{\today}
\maketitle

\tableofcontents

\section{Introduction}
Particles that carry a fraction of the electron charge, $Q=\epsilon e$, also called millicharged particles (mCPs) are an elegant extension to the standard model (SM). The discovery of their simplest iteration, without the presence of a dark photon, would be a violation of the charge quantization hypothesis. Alternatively, mCPs could also arise as charges quantized under a dark force which kinematically mixes with the SM photon \cite{Izaguirre:2015eya}. 

Millicharged particles have attracted significant interest and there have been many attempts to detect them.
Collider and beam-dump experiments have been performed to search for mCPs and null results have lead to strong limits on their parameter space in the MeV-TeV mass range. These include LEP~\cite{akers1995search}, the SLAC millicharge experiment~\cite{Prinz:1998ua}, neutrino experiments~\cite{Magill:2018tbb}, the Argoneut experiment~\cite{Acciarri:2019jly}, MilliQan pathfinder experiment at the LHC~\cite{Ball:2020dnx} and the BEBC beam dump experiment~\cite{Marocco:2020dqu}. A plethora of experiments have been proposed to improve these tests and search for still smaller charges~\cite{Ball:2016zrp,Berlin:2018bsc,Kelly:2018brz,Harnik:2019zee}. At lower masses, stringent limits arise from absence of anomalous emission in stellar environments~\cite{Davidson:2000hf,Chang:2018rso} at masses below an MeV. The prospect of mCPs making up some or all of the dark matter (DM) has also received a lot of interest over the years. There are robust predictions for their relic density from the early universe~\cite{Dvorkin:2019zdi,Creque-Sarbinowski:2019mcm}, as well as numerous ways to detect them~\cite{Knapen:2017ekk,Blanco:2019lrf,Essig:2019kfe,Berlin:2019uco,Kurinsky:2019pgb,Barak:2020fql,Griffin:2020lgd} depending on their mass $m_Q$ and charge. Subcomponent millicharge DM (mCDM) has also been invoked to explain several recent experimental anomalies~\cite{Barkana:2018qrx,Munoz:2018pzp,Liu:2019knx,Kurinsky:2020dpb,Bloch:2020uzh,Harnik:2020ugb}. 

Since mCPs interact with the SM particles through a massless mediator, their transfer cross-section can be large at small velocities. As a result, mCDM with charge large enough to scatter in the atmosphere and Earth overburden of direct detection experiments, rapidly loses its virial kinetic energy and thermalizes with the environment. When it eventually reaches a direct detection experiment, it does not possess enough energy to deposit in the detector and can not be observed at direct detection experiments~\cite{Emken:2019tni}. However, these mCPs which are now cooled to the ambient temperature, get trapped due to Earth's gravity and build up for the duration of the Earth's existence. This can lead to mCP densities on Earth up to fourteen orders of magnitude larger than that of the virial population in the galaxy \cite{Pospelov:2020ktu}. This slow, albeit dense population requires novel detection strategies, some of which include mCP particle-antiparticle annihilation in a large-volume detector~\cite{Pospelov:2020ktu} and accelerating mCPs present in electrostatic accelerator tubes to higher energies, sufficient for subsequent direct detection~\cite{Pospelov:2020ktu}. 
If the charge is large enough, negatively charged mCPs can bind with large positively charged SM nuclei, thereby creating fractional charge for a macroscopic material. Searches for such fractional charges bound to a sample material include Millikan-like oil-drop experiments~\cite{kim2007search}, as well as more recent levitation experiments with microspheres~\cite{moore2014search,afek2020limits}. However, these limits hinge critically on the assumption that the negatively charged mCPs bind to matter, thereby restricting their validity to only large charges and masses. 

In this work, we point out an alternate search strategy; the remarkably stable trapped ions developed for metrology and quantum information science are the ideal targets to detect an ambient thermalized mCP population. There is a long history of using trapped charged SM  particles for particle physics applications. Trapped ions have been used to measure the anomalous magnetic moment of the electron~\cite{hanneke2011cavity}, the electron's electric dipole moment~\cite{cairncross2017precision}, proton and antiproton magnetic moments \cite{schneider2017double,smorra2017parts}, as well as time variation in fundamental constants which can be induced by ultralight bosonic dark matter~\cite{Roberts_2020}. Ion traps are also the preeminent candidate for qubits to realize quantum computing \cite{haffner2008quantum}. 

For all of the applications listed above and especially for magnetic moment measurements and quantum computing, it is important to keep the ion trapped for sufficiently long duration without heating from the surroundings. In the last couple of decades, there has been remarkable progress in reduction of the measured heating rate of trapped ions \cite{hite2012100}. Further progress is expected and is an active area of research in order to achieve scalability of multi-ion systems and increased sampling rate in precision measurements. We briefly explain why the above properties make ion traps the ideal candidate for detecting mCPs next.

\subsection{Summary of Findings}
\label{sec:exec}

We propose the use of ion traps as detectors for millicharged particle (mCP) dark matter.  While an individual ion constitutes a much smaller target mass than any other dark matter direct detection experiment, a trapped ion has significant advantages for detection of mCPs including isolation from the environment, a lower energy threshold for detection, and a larger scattering cross section with mCPs.  These advantages outweigh the small target mass, allowing ion traps to reach many orders of magnitude past other detection methods for mCPs. 

Significant effort has been put into isolating trapped ions from their environments.
Thus trapped ions are now sensitive to small energy depositions down to $\sim$ neV.  
The dense, thermal gas mCPs, if they exist on Earth, would permeate the detector and can scatter off the ion.  
Depending on the nature of the experiment, it may be possible to measure individual scattering events or just an overall heating rate of the ion.
The mCPs are thermalized with the walls of the detector which is held at a temperature much higher than the temperature of the trapped ion. Thus, the higher-energy mCPs can transfer kinetic energy which can be detected either as a single jump of the trapped ion, or an accumulation of several scatters resulting in heating of the ion.  We will discuss both types of signals.
The high degree of isolation of the ion achievable in these traps makes them sensitive to scattering rates for mCDM over a wide range of parameter space.

Ion traps also make excellent mCDM detectors because of their low energy thresholds that are set by the energy-level spacings in the trap that can be as low as $\sim \text{neV}$.  This allows detection of single scattering events with this energy and also allows such low-energy scatters to contribute to the heating rate.

Furthermore, because they are charged, ions make excellent targets for mCP scattering.  mCPs scatter with ions via Rutherford scattering which is greatly enhanced at small relative velocities and momentum transfers.  Since the mCPs are thermalized with the walls, they generally have much lower velocities than virialized dark matter. The corresponding boost to the scattering cross section combined with the large number density of the thermalized mCPs makes direct detection with single ions viable.

The rest of this paper is organized as follows.
In Sec.\,\ref{sec:traps} we provide a description of ion traps. In Sec.\,\ref{sec:mcpdyn}, we provide an overview of the mCP density on Earth, as well as their dynamics, including propagation through the trap. The detection signals are explained in Sec.\,\ref{sec:signals} and results and projections are presented in Sec.\,\ref{sec:results}. We conclude with a discussion in Sec.\,\ref{sec:discussion}. 

\section{Ion traps}
\label{sec:traps}
\begin{figure}[h!]
\centering

    \centering
    \includegraphics[width=\linewidth]{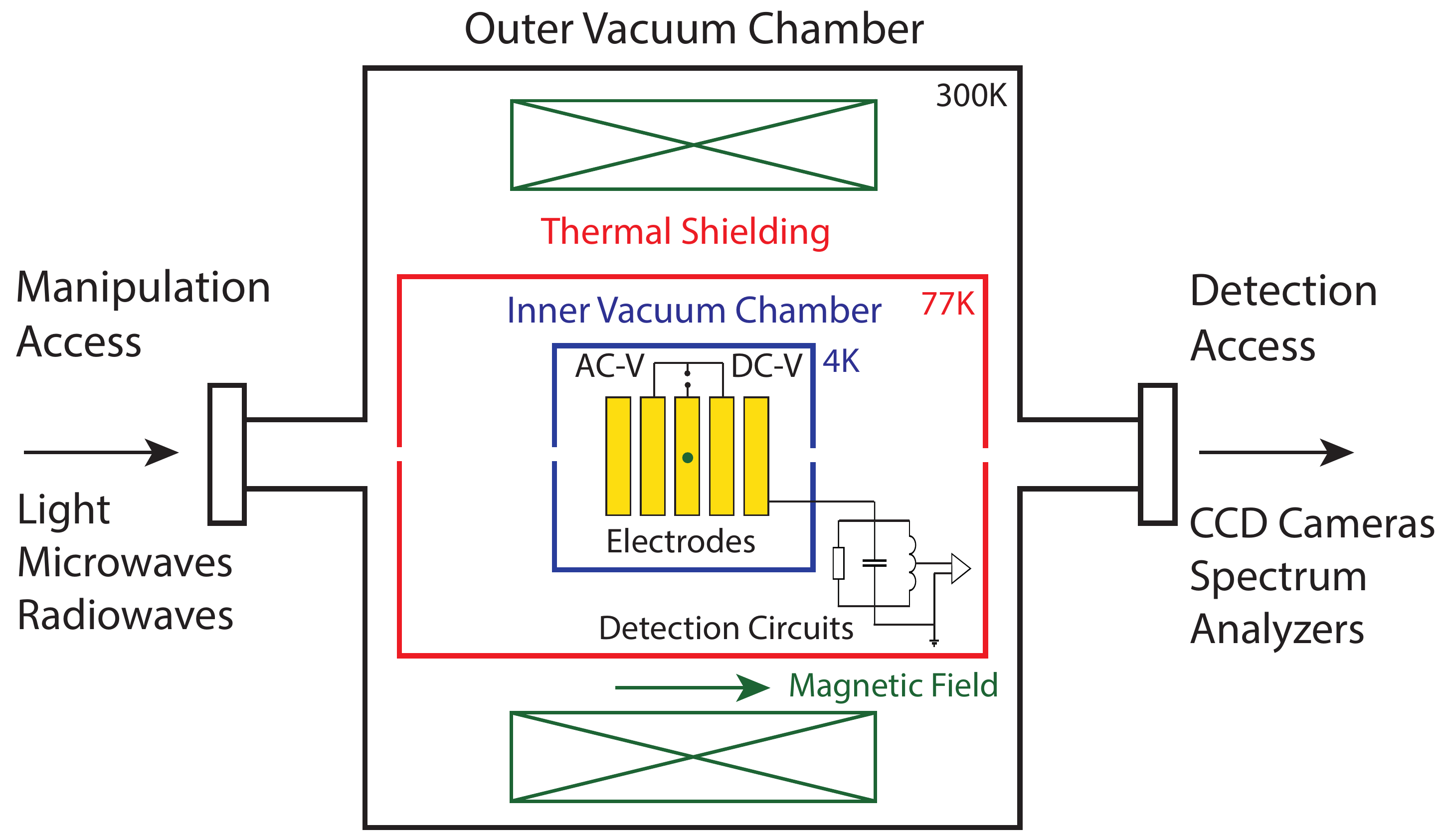}

    \caption{Elements common to typical trap experiments. A set of electrodes supplied by AD and DC voltages is mounted in a vacuum chamber, optionally in a strong magnetic field. Cryogenic traps have in addition thermal shielding and cryogenic vacuum chambers at low temperatures in their surroundings. The particles are manipulated with laser beams, microwave and radio-frequency generators. Signals are read-out by detecting image currents or fluorescent light. }
    \label{fig:Trap}
\end{figure}

Traps for (single) charged particles belong to the basic tool-set of atomic, molecular, and optical physics \cite{safronova2018search}. They have wide applications in determinations of atomic masses and fundamental constants,  in precision measurements to test fundamental symmetries, and in quantum information technology. 
Typical ingredients common to all trap experiments are shown in Fig$.\,$\ref{fig:Trap}. The core of the experiments are usually sets of electrodes supplied by AC and DC voltages, optionally placed in strong magnetic fields.  
The electrodes are mounted in vacuum chambers, in some cryogenic trap experiments pressures on the level of 10$^{-18}\,$mbar are achieved,  which provides ultra-long particle storage times \cite{sellner2017improved} and enables non-destructive long-term studies at low background. Thermal shielding and an additional insulation vacuum chamber usually surround the inner vacuum chamber. The trapped particles are manipulated, cooled and excited via laser, microwave and radio-frequency-drives. The experimental signals for ultra-sensitive precision studies, frequency measurements, and monitoring of quantum information-processing protocols are either image-currents picked up by sensitive detection circuits, or fluorescence signals. The signals are acquired and processed with spectrum analyzers and charge coupled device (CCD) cameras, respectively.

\begin{table*}
\begin{center}
\begin{tabular}{ ||c|c|c|c|c|c|c|c| }
\hline
 Experiment & Type & Ion & $V_z$  &  $T_{\rm wall}$ & $\omega_p$ [neV] & $T_{\rm ion}$[neV] & Heating Rate (neV/s) \\ 
 \hline 
 Hite et al, 2012 \cite{hite2012100}& Paul &$^9\text{Be}^+$ & 0.1 V&300 K&$\omega_z=14.8$ &$14.8$&640 \\
 Goodwin et al, 2016\,\cite{goodwin2016resolved} &Penning&$^{40}\text{Ca}^+$ & 175\,V  &300\,K& $\omega_z=1.24$ & $1.24$ & 0.37 \\
 Borchert et al, 2019\,\cite{borchert2019measurement}&Penning&$\bar{p}$& 0.633\,V& 5.6\,K & $\omega_+=77.4$&7240 & 0.13 \\
 &&& & & $\omega_-=0.050$& &  \\ \hline
\end{tabular}
\end{center}
\caption{List of ion traps and the relevant experimental parameters used for setting limits in this paper. The ion used, $V_z$,  potential barrier in the axial direction and $T_{\rm wall}$, the temperature of the walls of the trap and $\omega_p$, the fundamental frequency of the trap in the relevant direction are listed. Also listed are $T_{\rm ion}$, the temperature of the ion in the trap and the measured heating rate. }
\label{table:data}
\end{table*}

\subsection{Penning Traps}
In a Penning trap, there is a strong magnetic field $B_0$ superimposed with an electrostatic quadrupolar potential $\Phi(z,\rho)$, which is attractive along the magnetic field axis. The motion of a charged particle in such crossed static fields is composed of harmonic oscillator modes of three independent types. The modified cyclotron and the magnetron modes correspond to oscillations perpendicular to the axis, while the axial mode corresponds to oscillations along the magnetic field lines. Associated with the mode oscillations are the three trap eigenfrequencies $\nu_+$, $\nu_-$, and $\nu_z$, respectively. Room-temperature traps such as that in Ref.\,\cite{goodwin2016resolved}, cool and trap ions such as $^{40}\textrm{Ca}^+$ or $^{9}\textrm{Be}^+$ to the axial ground state using optical sideband cooling~\cite{mavadia2014optical}. A delay period after cooling is followed by subsequent spectroscopy which determines the final state of the ion, thus measuring the heating rate. The lowest heating rate achieved thus far at room temperature is reported in Ref.\,\cite{goodwin2016resolved} where the increase in the number of phonons of the axial mode was reported to be $\dot{n}=0.3/\textrm{s}$ with the axial frequency $\nu_z=0.3\,\textrm{MHz}$, see  Tab.\,~\ref{table:data}. \\
Particularly interesting for the detection of mode energy changes are experiments dedicated to direct measurements of nuclear magnetic moments such as those of the proton \cite{schneider2017300}, the antiproton \cite{smorra2017parts}, or $^3$He$^{2+}$ \cite{mooser2018new}. These experiments operate advanced cryogenic Penning-trap systems consisting of multi-trap assemblies, common to all of them is a so-called analysis trap with a strong superimposed magnetic inhomogeneity $B(z)=B_0+B_2 z^2$, where $B_2$ characterizes its strength \cite{ulmer2011observation}. 
The interaction of the magnetic bottle with the particle's magnetic moment $\mu_z=\mu_++\mu_-+\mu_s$ results in a magnetostatic axial energy $E_{B,z}=-\mu_z B_z$, where $\mu_+$ and $\mu_-$ are the orbital angular magnetic moments associated with the modified cyclotron and the magnetron mode, while $\mu_s$ is the spin magnetic moment. 
As a result, the particle's axial frequency $\nu_z=\nu_{z,0}+\Delta\nu_z (n_+,n_-,m_s)$ becomes a function of the radial trap eigenstates $n_+$ and $n_-$, as well as the spin eigenstate $m_s$ with
\begin{align}
\Delta\nu_z=\frac{h\nu_+}{4\pi^2 m \nu_z}\frac{B_2}{B_0} \left(\left(n_++\frac{1}{2}\right)+\frac{\nu_-}{\nu_+}  \left(n_-+\frac{1}{2}\right)+\frac{g}{2} m_s\right).
\end{align}
Measurements of single-particle magnetic moments in Penning traps rely on the detection of axial frequency shifts $\Delta\nu_{z,\text{SF}}$  induced by driven spin quantum transitions $\Delta m_s=1$. Since nuclear magnetic moments are about three orders of magnitude smaller than the Bohr magneton, these experiments require highest sensitivity with respect to magnetic moments, which is usually achieved by the utilization of magnetic bottle strengths of the order $B_2\approx100\,$kT/m$^2$ to $B_2\approx400\,$kT/m$^2$. Combined with continuous measurements of the axial frequency, such strong magnetic bottles provide excellent resolution of the radial mode energies with 
\begin{eqnarray}
\frac{(\Delta\nu_z)}{\Delta E_\rho}=\frac{1}{4\pi^2m \nu_z}\frac{B_2}{B_0}\approx 1 \frac{\text{Hz}}{\mu \textrm{eV}}, 
\end{eqnarray}
while in axial frequency measurements resolutions of order  200$\,$mHz$\cdot\sqrt{\text{s}}/\sqrt{t_\text{avg}}$ are achieved, $t_\text{avg}$ being the averaging time which is typically on the order of several tens of seconds. Transition rates in the radial modes
$(dn_{+,-})/dt\propto(n_{+,-}/\omega_{+,-} ) S_E (\omega_{+,-})$ lead to random walks in radial energy space and to axial frequency diffusion. Here  $S_E (\omega_{+,-})$ is the power spectral density of a noisy background drive, and $n_{+,-}$ is the principal quantum number of the modified cyclotron ($n_+$) / magnetron ($n_-$) oscillator. The scaling of the heating rate with $n_{+,-}$ is related to eigenstate-overlap of harmonic oscillator states \cite{ulmer2011observation}. \\ 
By analyzing time sequences of axial frequency measurements $\nu_z (t)$, the average radial quantum transition rates are obtained. With a highly optimized trap setup with which the antiproton magnetic moment was measured with 1.5 parts per billion precision, the BASE experiment at CERN reports on the observation of absolute cyclotron transition rates of 6(1) quanta per hour~\cite{borchert2019measurement}. Together with the determination of the $n_+$ state during the recorded measurement, this result is  consistent with a projected ground state heating rate of 0.1 cyclotron quantum transitions per hour, setting an upper limit which is by a factor of 1800 lower than the best reported Paul-trap heating rates, and by a factor of 230 lower than the best room-temperature Penning trap. These numbers are summarized in Tab.\,\ref{table:data}. Note that the antiproton experiments are conducted in a background vacuum of $\approx 10^{-18}\,$mbar \cite{sellner2017improved}, constraining parasitic heating induced by collisions with  background-gas to a level of $4\times10^{-9}$/s.

\subsection{Paul Traps}
Paul traps or radiofrequency traps utilize an oscillating voltage to confine in the perpendicular direction instead of the magnetic field used for the same purpose in the Penning trap. Paul traps have a rich history of being used as mass spectrometers and more recently in building quantum computers~\cite{brownnutt2015ion}. 

The effective potential in the presence of both DC and AC potentials can be written as
\begin{equation}
    \psi(x,y,z,t)=\left(U_{\rm DC}+V_{\rm AC} \cos \Omega t\right)\frac{r^2+2 (z_0^2-z^2)}{r_0^2+2z_0^2} .
\end{equation}
Here $z$ is the axial direction, $r$ the radial direction with the distance to the electrodes given by $r_0$ and $z_0$. The rapidly oscillating potential creates a pseudopotential for charges of both signs and this leads to approximately simple harmonic motion very close to the trap center.  After laser cooling to the ground state, the total heating rate can be measured via Raman sideband technique~\cite{turchette2000heating}. There has been extensive study of the heating rate in Paul traps and its dependence on distance to electrodes, wall temperature, trap temperature~\cite{brownnutt2015ion} as well as ion beam treatment of electrodes~\cite{hite2012100}. Electric field noise from the electrodes has been identified as the dominant heating source, with the dependence on distance scaling as $d^{-2}$~\cite{brownnutt2015ion} to $d^{-4}$~\cite{daniilidis2011fabrication}. Although heating rates are lower for bigger traps~\cite{poulsen2012efficient}, smaller sized traps employ shallower potential wells $\approx0.1~\textrm{V}$. As we shall see, this allows mCPs with large charge to reach the trap and hence provide complementary reach at large charge parameter space. Hence we reinterpret limits only from a microtrap~\cite{hite2012100}.  The heating rates reported in ~\cite{hite2012100} are $\dot{n}=43/\textrm{s}$ for the axial frequency $\nu_z=3.6~\textrm{MHz}$. These numbers are tabulated in Tab.\,~\ref{table:data}.

\section{Millicharged Particle Dynamics}
\label{sec:mcpdyn}
\subsection{Terrestrial Accumulation} 
If mCDM exists and is virialized in the galaxy, there is a non-zero flux of mCPs flowing through the Earth at all times.  This mCDM stops in the atmosphere or rock overburden for large enough charge, and can accumulate on Earth. This process was treated in detail in~\cite{Emken:2019tni} and the subsequent accumulation in~\cite{Pospelov:2020ktu}. We provide here a summary of the relevant results of these papers that are used in Section.\,\ref{sec:results} and refer the reader to ~\cite{Pospelov:2020ktu} for details.

Following~\cite{Pospelov:2020ktu} we consider effectively millicharge particles mediated by a dark photon which kinematically mixes with the SM photon. The dark photon mass is taken to be large enough ($m_{A'} \gtrsim 10^{-12}~\textrm{eV}$) such that the effect of large-scale electric and magnetic fields can be ignored while considering mCP propagation\footnote{We leave to upcoming work the calculation of the accumulation of millicharged dark matter in cases where the electric and magnetic fields are relevant \cite{blprfuture}. Once such calculations are complete, our given limits on the number density in the lab can be translated to limits on dark matter fraction in theses cases.}.  In this limit, the mCPs with mass $m_Q\ge 1$\,GeV are stuck on the Earth after thermalization.\footnote{Millicharged particles with masses below 1\,GeV can accumulate on Earth temporarily before evaporating. We leave limits on these masses for future work.} The volume averaged DM number density on Earth, $\langle n_Q \rangle$ can be several orders of magnitude larger than the virial DM density. It is given by, 
\begin{align}
\label{start}
\langle n_Q \rangle&= \frac{\pi R_{\oplus}^2 v_{\rm vir}  t_{\oplus}}{4/3 \pi R_{\oplus}^3}n_{\rm vir} \nonumber \\
&\approx \frac{3\times 10^{15} }{\textrm{cm}^3} \frac{t_{\oplus}}{10^{10}\,\textrm{y}} f_Q \frac{\rm GeV}{m_Q},
\end{align}
where, $R_{\oplus}$ and $t_{\oplus}$ are the radius and age of the Earth and $n_{\rm vir}$ and $v_{\rm vir}$ are the galactic virial number density and velocity. However, the equilibrium density profile is peaked at the Earth core. This density $n_{\rm static}$ was calculated by taking into account the Earth's temperature and density variations in \cite{Neufeld:2018slx}. However, the sinking to the Earth's core is not immediate, and there is a dynamic population, the so-called ``traffic-jam" density, $n_{\rm tj}$, given by,
\begin{equation}
    n_{\rm tj}=n_{\rm vir}\frac{v_{\rm vir}}{v_{\rm term}} .
\end{equation}
Here, $v_{\rm term}$ is the terminal velocity in rock, given in \cite{Pospelov:2020ktu}. Finally, the density in the laboratory, $n_{\rm lab}$ is given by, 
\begin{equation}
n_{\rm lab} = \textrm{Max}\left( n_{\rm static}, \textrm{Min}\left( n_{\rm tj} , \langle n_Q \rangle \right)\right)
\label{eq:regimes}
\end{equation}

These results are applicable only for $\epsilon$ large enough, such that it stops in the corresponding overburden. According to \cite{Pospelov:2020ktu} this is valid only for
\begin{align}
\epsilon&\gtrsim 2\times10^{-4} \sqrt{\frac{m_Q}{\rm GeV}} \quad &&\textrm{surface,} \nonumber \\
&\gtrsim 3\times 10^{-6}\sqrt{\frac{m_Q}{\rm GeV}} \quad &&\textrm{1 km mine.} 
\label{eq:applicable}
\end{align}

It is also important to comment on the asymmetry of the mCP population. If the mCDM is a symmetric population with equal number of particles and antiparticles, accumulation on Earth can result in Sommerfeld-enhanced annihilations which prevent build up. We consider here the asymmetric case such that opposite charges are carried by different species just like the the SM proton and electron, such that annihilations are absent. In this scenario, for large enough $\epsilon$, the negatively charged mCPs can form deep bound states with the large positively nuclei, whereas the positive mCPs can only bind with the less massive electrons which also have smaller charge compared to the heavy SM nuclei. In ref.~\cite{Pospelov:2020ktu} it was pointed out that positive mCPs with $\epsilon \ge 0.042$  bind to electrons at 300 K (room temperature). We find that such bound states are temporary with electrons rapidly preferring to bind with ions which possess larger charge than the positive mCPs. Thus the terrestrial population of positive mCPs remains free of binding for $\epsilon\lesssim 1$ and is present as a locally thermalized population that is diffusing everywhere. Thus, the results we derive apply to all positive charges with $\epsilon \lesssim 1$ and negative charges which do not bind with nuclei, i.e \cite{Pospelov:2020ktu} $\epsilon < \frac{m_e}{\mu_{Q,N}}$ where $\mu_{Q,N}$ is the reduced mass of the mCDM-nuclear system.

It is important to emphasize that the limits put on the ambient number density are applicable to mCPs charged directly under the SM photon as well as to ones mediated by a dark photon as long as the dark photon mass is below the relevant momentum transfer for scattering with ions which is around 1 eV.   

\subsection{Passage Through Apparatus}

\label{trappassage}
We next turn to the trajectory of mCPs through the trap peripherals in order to reach the trapped ion.  We need to know the density and temperature of the mCPs reaching the ion.  In this subsection we explain the main factors entering the calculation, but as the calculation itself is somewhat involved we leave the details to Appendix \ref{Appendix:passage}.  Our main result is in Eq.\,\eqref{eqn:qion} relating the number density of mCPs at the position of the ion, $n^Q_{\rm ion}$, to the ambient number density on the Earth $n_{\rm lab}$.  
This is the equation we use to set our limits on mCPs.

At equilibrium, mCPs are expected to have roughly uniform density near the Earth's surface, including permeating all materials.  However the conditions of the experiment can affect this naive expectation for several reasons.

First, some of the ion traps we consider are cryogenic.  Over essentially all of our parameter space the mCPs have a short interaction length in material and so will rapidly thermalize to the cryostat temperature as they enter the experiment.  By itself this would lead to an increase in mCP density by a factor linear in mCP velocity ($\propto \sqrt{T}$) because the fluxes entering and leaving the cryostat must be equal in equilibrium.

Second, the ion traps are surrounded by metal which has a work function that can affect the passage of mCPs.  This is only relevant for mCPs of relatively large charge ($\epsilon \gtrsim 10^{-2}$) but for those it can be a significant effect.  The work function for mCPs, which we will call $\epsilon \phi$, is not simply $\epsilon$ times the work function for electrons.  For electrons the work function arises from several contributions of varying signs including e.g.~the binding to the lattice of nuclei, the Fermi sea of other electrons, and surface effects such as the ``double layer" or the image charge potential.  Several of these do not apply or are negligible for mCPs.  Recall we are only considering positively charged mCPs since the negative ones may be stuck deeply bound to some nucleus somewhere on the Earth.  Thus the main effects are repulsion by the double layer and possibly also repulsion from the nuclei.   We consider the work function for mCPs in more detail in  Appendix \ref{Appendix:work function}.  Our conclusion is that the work function is repulsive for positively charged mCPs and thus every metal sheet provides a barrier for mCPs to cross. 
The size of the potential barrier that has to be crossed is $\phi \sim \text{few} \, \eV$.
Note that for an experiment at room temperature ($T \sim 0.03 \, \eV$) the metal barriers are then irrelevant for charges $\epsilon \lesssim 10^{-1}$ because the Boltzmann tail easily pushes a fast enough rate of mCPs over the barrier.  For a cryogenic experiment at $T \sim 6 \, \text{K} \sim 5 \times 10^{-4} \, \eV$, the metal barrier will be relevant for charges $\epsilon \gtrsim 10^{-3}$ and essentially insurmountable for charges $\epsilon \gtrsim 10^{-2}$.  As we show in Appendix \ref{Appendix:passage}, the most important effect comes from an experiment encased in two different metals where the work function for mCPs rises from the outer metal to the inner metal.  We will take this difference to be $\Delta \phi = 3 \, \eV$ for all experiments we consider since this will be a conservative estimate as we show in Appendix \ref{Appendix:work function}.

Third, the ions are always in a region of ultrahigh vacuum.  This means that pumps were used to remove the Standard Model (SM) particles.  Given a short interaction length of the mCP in materials, these pumps could remove the mCPs from the ion chamber as well.  In one of the experiments we consider (Goodwin et al.~\cite{goodwin2016resolved}) this effect is not relevant because the trap is at room temperature and the region of sensitivity is at low enough $\epsilon$ that the millicharges pass easily through the walls\footnote{The sensitivity is limited to be below $10^{-3}$ charge because of the large trap potential as will be discussed in Section \ref{sec:results}.}.  In the other two experiments we consider (Hite et al.~\cite{hite2012100} and Borchert et al.~\cite{borchert2019measurement}), the vacuum pumps are turned off well before the actual data taking is begun.  In the case of Borchert this is at least a year, while for Hite we conservatively assume it is only a day\footnote{The final answer is only logarithmically sensitive to this timescale anyway.}.  And of course mCPs are always continually flowing in from the walls of the vacuum chamber.  This would rapidly refill the trap region and this effect would not be relevant, except for the largest charges where the refill can be slow because of the work function of the surrounding metal.  This does mean that, depending on the parameters of the mCP and of the ion trap, the number density in the trap may be either in the equilibrium regime or in the filling regime.  This is why Eq.\,\eqref{eqn:final trap density} has two different regimes.  Eq.\,\eqref{eqn:final trap density} relates the number density of mCPs inside the trap $n_{\rm trap}$ to the ambient number density on the Earth $n_{\rm lab}$.  There is then one remaining step to find the number density of mCPs at the position of the ion.

Fourth, the ion trap itself has applied electromagnetic fields to trap the ion.  These can affect the passage of the mCPs, though again this is only relevant for larger charges $\epsilon \gtrsim 10^{-3}$.
Tab.\,\ref{table:data} lists parameters of the various experiments we consider.  All the experiments use an electric DC potential to confine the ion in the axial direction.  The height of the potential barrier along the axial direction is listed as $V_z$.  The mCPs in the trap are thermalized to the temperature of the walls of the trap $T_{\rm wall}$.  Starting from the number density of mCPs calculated inside the trap $n_{\rm trap}$ in Eq.\,\eqref{eqn:final trap density}, we take a Boltzmann suppression on the number density which can make it up the axial barrier height.  Thus we take the final number density at the position of the ion to be given by Eq.\,\eqref{eqn:qion}.  For the experiments we consider this Boltzmann suppression is relevant for setting the ceiling (largest $\epsilon$ values) of the Goodwin regions in Figure \ref{fig:evsm}, but is irrelevant for the other regions. 
The Penning traps use a magnetic field to confine the ions in the radial direction.  This B-field will cause mCPs with a large enough charge to circle around the axial B-field lines but they are still free to move along the axis.  The magnetic field does not change the phase-space density of the mCPs and so it does not significantly affect our signal.  The Paul traps use an RF potential to confine the ion in the radial direction.  This potential around the minimum is locally attractive for the positive mCP independent of the sign of the charge of the trapped ion.  An mCP coming from far outside the RF fields in a radial direction could give a barrier in principle.  However for mCPs approaching along the axial direction, the RF pseudopotential in the axial direction is very weak and negligible.  And then  such mCPs will actually be concentrated in the radial direction towards the ion at the center of the trap since the potential is locally attractive.  We conservatively ignore this possible (Sommerfeld-like) enhancement though it could be large. For large enough mCP masses, $m_Q \gtrsim 10^{10} \textrm{GeV}$, the free-fall under gravity 
can generate velocities much larger than the thermal velocities assumed. This can increase the heating rate further. We conservatively ignore this effect and leave its consideration for future work.

\section{Observables for Millicharged Particles in Ion Traps}
\label{sec:signals}
In this section we consider the interaction of ambient mCPs with ion traps to identify observables for detection. As seen in Section.~\ref{trappassage}, the mCPs enter the ion trap with effective temperature $T_Q \ge T_{\rm wall}$, where $T_{\rm wall}$ is the wall temperature. $T_{\rm wall}= 5.6$ K for the cryogenic trap we consider~\cite{borchert2019measurement} and $T_{\rm wall}=T_{\rm room}\approx 300$ K for room temperature traps. The ions in the trap are at a much colder temperature $T_{\rm ion} \ll T_{\rm wall}$. With this hierarchy of temperatures, the mCPs can cause two types of signal.

The first signal involves individual scattering events that impart energy $E_{\rm ion}$ to the ion thus leading to a change in its harmonic oscillator quantum number. This signal is very similar to dark matter scattering in a conventional dark matter detector. The rate of events has to be slower than the rate at which the ions are interrogated. $E_{\rm ion}$ also needs to exceed the energy resolution $E_{\rm res}$ for detection. 

The second type of signal is the heating of the trapped ion due to collisions with multiple mCPs. In this case, the individual hits $E_{\rm ion}$ can be smaller than $E_{\rm res}$ and only the sum needs to exceed this resolution. 

For both of these signal types, we only consider individual energy transfers $E_{\rm ion}$ much larger than the typical energy spacing of the trap $\omega$, for concreteness $E_{\rm ion} \ge 10\omega$. \footnote{In principle, energy transfers $E_{\rm ion} =n\times \omega$ where all integers $n\ge 1$ are allowed. We restrict $E_{\rm ion} \ge 10 \times \omega$, a conservative choice that helps avoid form-factor calculations.} In this limit, the trapped ion can be approximated as a free particle with initial energy $T_{\rm ion}$ and final energy $T_{\rm ion}+E_{\rm ion}$. Equivalently, for energy transfers much larger than the spacing, the form-factor that incorporates wave-function overlap can be approximated to unity.  
We start quantifying both of these signals by the angular differential cross-section given by the Rutherford formula,
\begin{equation}
    \frac{d\sigma}{d\Omega}=\frac{2\pi\alpha^2 \epsilon^2}{\mu^2 v_{\rm rel}^4\left(1-\cos\theta\right)^2} ,
\end{equation}
where $\theta$ is the scattering angle. We next introduce kinematic variables that  simplify the computation of the scattering rate. The results are presented here, with the details of the derivation presented in Appendix.~\ref{app1}. The incoming mCP and trapped ion velocities are assumed to be $\mathbf{v_Q}$ and $\mathbf{v_{\rm ion}}$ respectively. 
The center of mass (CM) velocity $\mathbf{v_{CM}}$ is given by,
\begin{equation}
    \mathbf{v_{CM}}=\frac{\left(m_{\rm ion} \mathbf{v_{\rm ion}}+m_Q \mathbf{v_Q}\right)}{m_{\rm ion}+m_Q} .
\end{equation}

The change in velocity of the ion, $\mathbf{\Delta v_{\rm ion}}$ is given by,
\begin{align}
    \mathbf{\Delta v_{\rm ion}}=\frac{m_Q}{m_{\rm ion}+m_Q}
    \left[\left(\cos \theta-1 \right)
    \left(\mathbf{v_{\rm ion}}-\mathbf{v_Q}\right)
    \right. 
    \nonumber \\ 
    \left.
    +\sin \theta |\mathbf{v_{\rm ion}}-\mathbf{v_Q}|
    \mathbf{n_{\perp}}
    \right] .
\end{align}
The transferred energy $E_{\rm ion}$ is given by,
\begin{equation}
    E_{\rm ion} = m_{\rm ion} \mathbf{v_{CM}}.\mathbf{\Delta v_{\rm ion}} .
\end{equation}

Given a threshold $E_{\rm thr}$, the single event rate $R_{\rm single}$ with energy transfer $E_{\rm ion}$ above this threshold is,
\begin{align}
    R_{\rm single}\left(E_{\rm ion}\ge E_{\rm thr}\right)=n^Q_{\rm ion} \int d^3 \mathbf{v_Q} g_Q  \int d^3 \mathbf{v_{\rm ion}} g_{\rm ion}  \int d \Omega  \nonumber \\ |v_Q-v_{\rm ion}| \frac{d\sigma} {d \Omega} \Theta \left(|E_{\rm ion}|-E_{\rm thr}\right) . 
\end{align}
Here $n^Q_{\rm ion} $ is the number density of mCPs at the ion position and is given by Eq.\,\eqref{eqn:qion}, $g_{\textrm{ion}(Q)}$ is the Maxwell--Boltzmann distribution for the ion/mCP with temperatures $T_{\rm ion}$ and $T_{\rm wall}$, respectively.

The heating rate per ion, $\dot{H}$ can be computed through,
\begin{align}
    \dot{H}&=n^Q_{\rm ion}  \int d^3 \mathbf{v_Q} g_Q  \int d^3 \mathbf{v_{\rm ion}} g_{\rm ion}  \int d \Omega |v_Q-v_{\rm ion}|  \nonumber \\ &\frac{d\sigma} {d \Omega}E_{\rm ion} \Theta \left(|E_{\rm ion}|-E_{\rm thr}\right) \Theta\left(E_{\rm samp}-|E_{\rm ion}|\right) . 
\end{align}
The Heaviside theta function ensures the inequality $E_{\rm thr}\le E_{\rm ion} \le E_{\rm samp}$. Here $E_{\rm samp}$ is defined so as to prevent the average heating rate from including contribution from extremely rare events. It is defined through,
\begin{equation}
R_{\rm single}\left(E_{\rm ion}\ge E_{\rm samp}\right) t_{\rm obs}=1 ,
\end{equation}
where $t_{\rm obs}$ is the total observation time.

 \section{Results}
 \label{sec:results}

\begin{figure*}
\centering
\begin{subfigure}{0.48\textwidth}
    \centering
    \includegraphics[width=\linewidth]{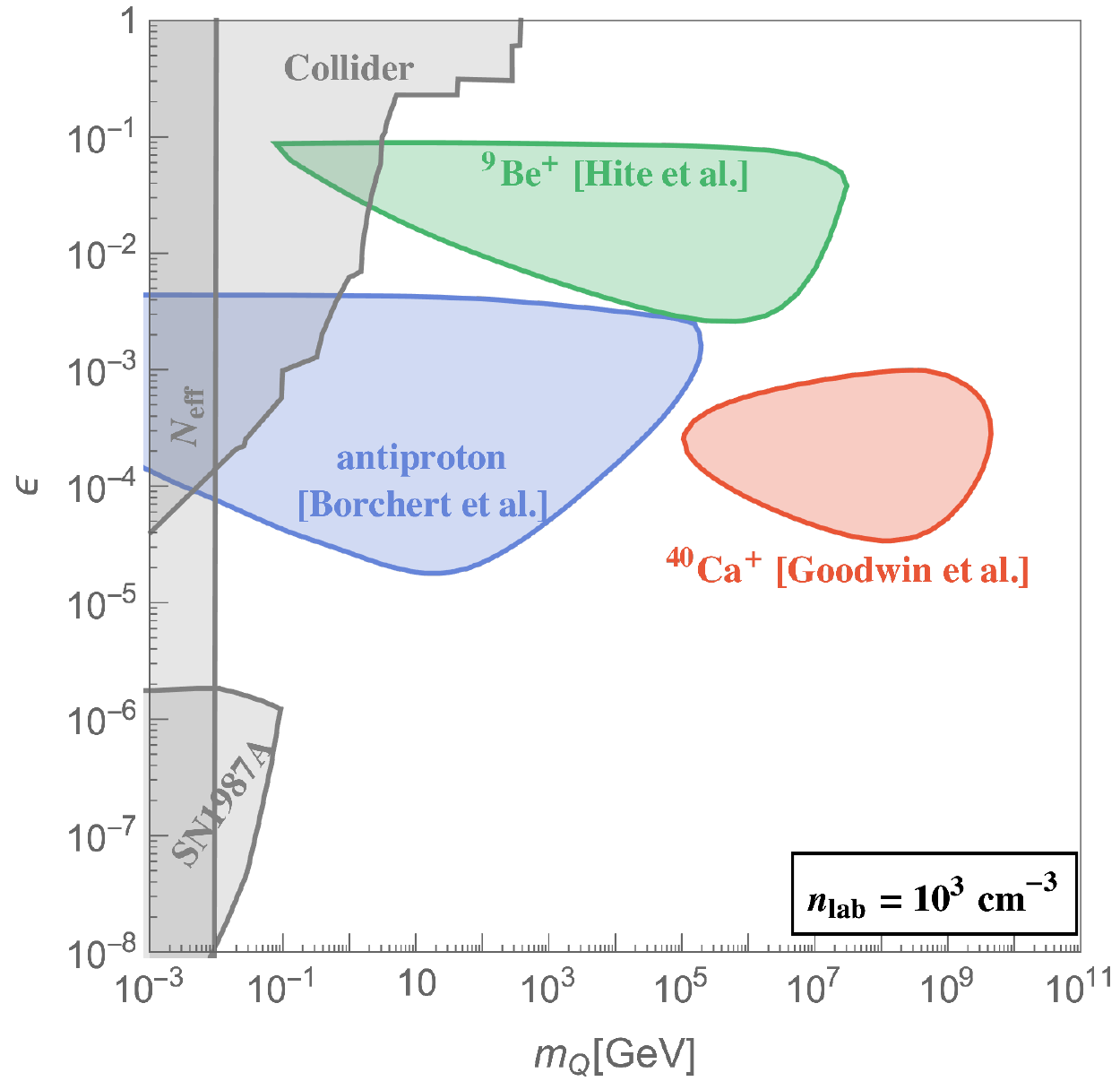} 

\end{subfigure}
\begin{subfigure}{0.48\textwidth}
    \centering
    \includegraphics[width=\linewidth]{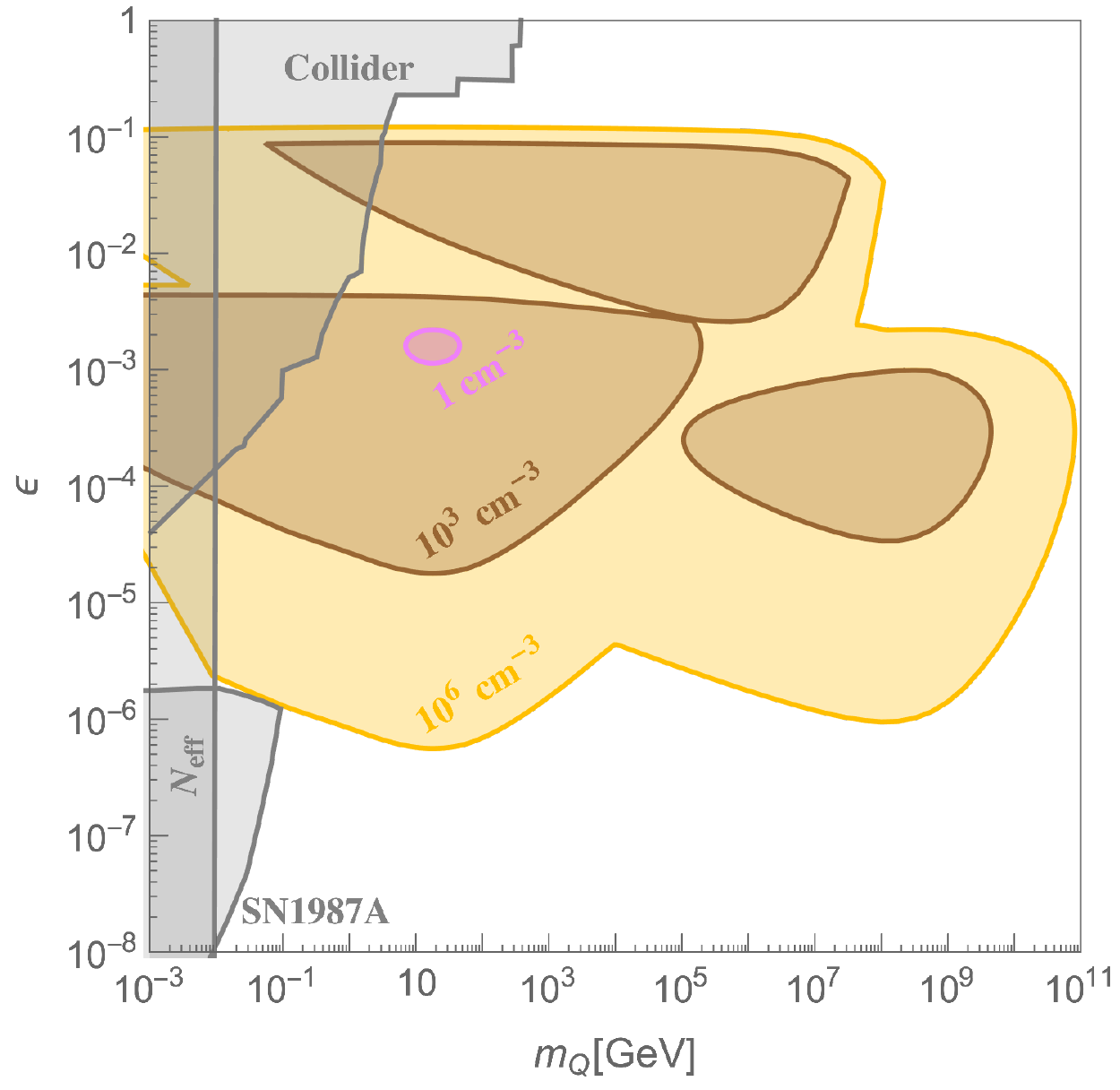} 
 
\end{subfigure}
    \caption{Compilation of new limits using existing heating measurements from various traps in Table \ref{table:data}: a room temperature Paul trap, Hite et al.~\cite{hite2012100}, a room temperature Penning trap, Goodwin et al.~\cite{goodwin2016resolved} and a cryogenic penning trap, Borchert et al. \cite{borchert2019measurement}. {\bf (left)} Comparison between traps for an ambient density $n_{\rm lab}=10^3 \textrm{cm}^{-3}$; {\bf (right)} Combined limits from the three traps for different  $n_{\rm lab}$} 
    \label{fig:evsm}
\end{figure*}
 In this section, we set constraints on mCPs and make projections for the future by using the expressions for the signal rates derived in Sec.\,\ref{sec:signals}. 
 
 \subsection{Limits from existing measurements}
 To obtain existing limits, we use the data presented in Tab.\,\ref{table:data}. All of the trap parameters for the $^9\text{Be}^+$  trap are taken from \cite{hite2012100} while the trap depth is conservatively taken to be $V_z=0.1$~V, an order of magnitude larger than the typical potential depths in microtraps\footnote{see for e.g.~\cite{warring2013techniques} from the same group where the potential depth is report to be $V_z$=5 mV}. While the rest of the parameters for the $^{40}\textrm{Ca}^+$ experiment are provided in ~\cite{goodwin2016resolved}, we obtained the potential depth $V_z=175$~V from the authors. Finally, for ref.\,\cite{borchert2019measurement}, we use the parameters $T_{\rm wall}=5.6$~K and $V_z$=0.6~V. While the analysis in ~\cite{borchert2019measurement} dealt with measuring the cyclotron mode ($\omega_+)$, the observable heating rate is equivalently a limit on the magnetron mode ($\omega_-$) also. While individual jumps in $\omega_-$ are unobservable with existing precision, the frequency shift due to the heating rate accumulates. Since the Rutherford cross-section increases for smaller energy transfers, we use $\omega_-$ to convert existing heating limits into limits on mCPs. 

 We start by plotting existing limits in Fig.\ref{fig:evsm} in the millicharge $\epsilon$ vs mass $m_Q$ parameter space for contours of constant ambient density $n_{\rm lab}$. As mentioned earlier, for mCDM, the lab density is expected to be orders of magnitude larger than the virial density, i.e. $n_{\rm lab} \gg n_{\rm vir}$. In the left panel, limits arising from the antiproton trap \cite{borchert2019measurement} in blue, $^{40}\textrm{Ca}^+$~\cite{goodwin2016resolved} in red and $^{9}\textrm{Be}$ in green \cite{hite2012100} for an ambient density of $n_{\rm lab}=10^3/\textrm{cm}^3$. The dominant limits arise from the antiproton trap owing to its superior heating rate as seen in Tab.\,\ref{table:data}. However, since it is a cryogenic trap, the limits disappear at $\epsilon \approx 10^{-2}$ owing to the suppression arising from mCPs finding it increasingly difficult to penetrate metals due to their work function. As explained in Section.~\ref{sec:mcpdyn} and Appendix.~\ref{Appendix:passage}, this factor roughly scales as $e^{-\frac{\epsilon \Delta \phi}{T_{\rm wall}}}$ where $\Delta \phi$ is the difference in work functions between two adjacent metals. Hence this suppression is ameliorated for room temperature traps like the $^9\textrm{Be}^+$ trap \cite{hite2012100} in green which extends to $\epsilon\approx 0.1$. While the $^{40}\textrm{Ca}^+$ trap in red is also at room temperature, the trap is at a potential of 175~V and thus mCP charges above $\epsilon \approx 10^{-3}$ do not reach the ion. However, there is reach to higher mCP masses for this trap as a result of the ion being more massive. The right panel of Fig.~\ref{fig:evsm}
 corresponds to combined limits from these three experiments for different $n_{\rm lab}$.  
 
 \begin{figure}
\centering

    \includegraphics[width=\linewidth]{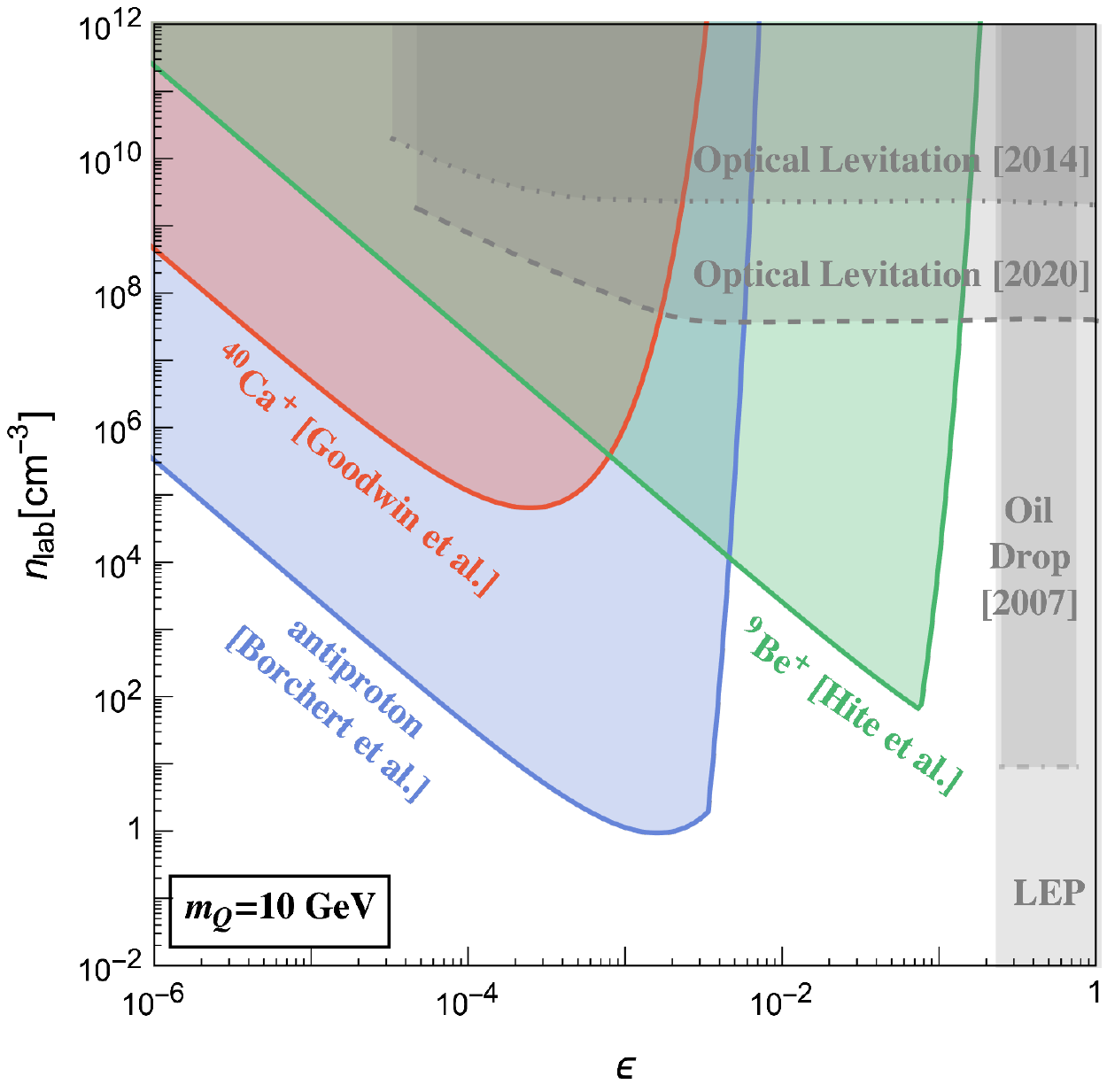} 

    \caption{
    Comparison of limits derived in this work from ~\cite{hite2012100,goodwin2016resolved,borchert2019measurement} with existing limits from Oil Drop~\cite{kim2007search}, levitation experiments~\cite{moore2014search,afek2020limits} and LEP~\cite{Marocco:2020dqu} for $m_Q=10$ GeV.}
    \label{fig:compexist}
\end{figure}

 As is clear from both figures, existing data for anomalous heating in traps sets exquisite bounds on mCPs thermalized locally. Bounds are applicable to orders of magnitude in the mCP mass $m_Q$ as well as many orders of magnitude in charge. Number densities as small as $n_{\rm lab}=1\, \textrm{cm}^{-3}$ are ruled out around the $\epsilon\approx 10^{-3}$ and $m_Q\approx 10\, \textrm{GeV}$ parameter point. 
 
In order to compare these limits on the ambient mCP population to ones that already exist in literature, we fix the mCP mass $m_Q=10$\,GeV and show limits in the $n_{\rm lab}$ vs $\epsilon $ plane in Fig.~\ref{fig:compexist}. The same color coding as Fig.~\ref{fig:evsm} is followed. In gray we show limits from LEP~\cite{Marocco:2020dqu}, as well as limits on mCPs bound in matter arising from Oil drop experiments~\cite{kim2007search}, and levitation experiments~\cite{moore2014search,afek2020limits}. As noted earlier, the limits on mCPs bound in matter are applicable only to negative mCPs with large enough charge such that binding with SM nuclei is possible. Furthermore, if mCP-SM bound states exist, there is no guarantee for these bound states to be evenly distributed all over the Earth. However, for the positive mCPs none of these caveats apply and they thermalize and distribute themselves over the entire Earth volume. Regardless, as seen in Fig.~\ref{fig:compexist} the limits obtained from ion traps are orders of magnitude stronger than the levitation experiments. For $\epsilon\approx 3\times 10^{-3}$, lab densities as small as $n_{\rm lab} \gtrsim 1 \textrm{cm}^{-3}$ are ruled out by the measured heating rate at the antiproton experiment \cite{borchert2019measurement}. 
 \begin{figure}
\centering

    \centering
    \includegraphics[width=\linewidth]{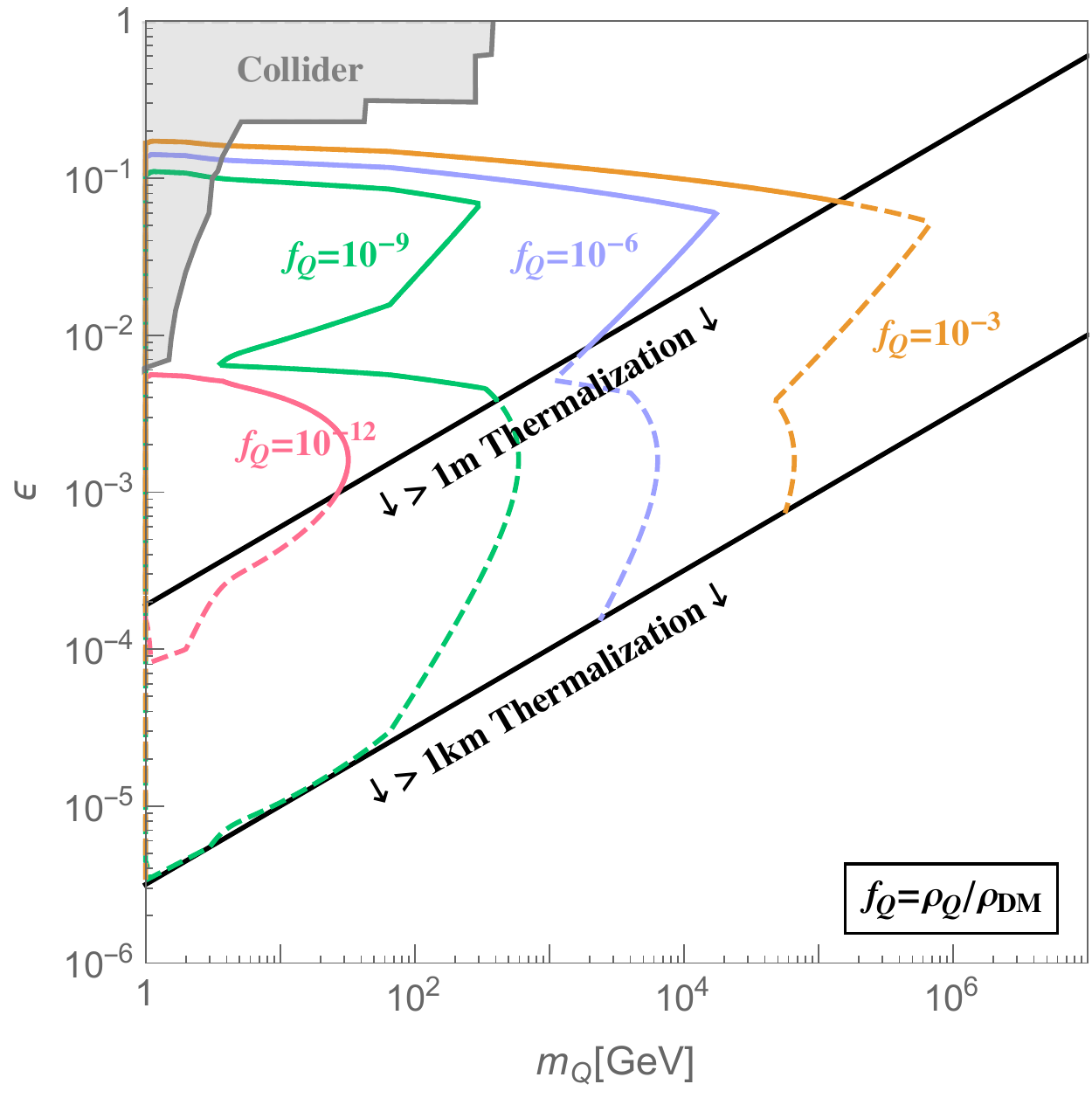}

    \caption{Limits and projections on the virial DM fraction in mCPs as a function of $\epsilon$ and $m_Q$. Since the current experiments were all conducted at the surface, the robust limits are only above the ``$>$1m Thermalization" line where the colored contours are solid. For a hypothetical deep mine experiment, the dashed part of the contours are also accessible.}
    \label{fig:virial}
\end{figure}

Next, in Fig.~\ref{fig:virial} we convert limits on $n_{\rm lab}$ into limits on the fraction of virial DM existing in mCPs, $f_Q= \frac{\rho_Q}{\rho_{\rm DM}}$. For this purpose we use $n_{\rm lab}$ from Eq.\,\ref{eq:regimes}. Existing limits from colliders are shown in gray. The solid parts of the colored contours correspond to the region where the incoming virial DM gets thermalized within 1 meter and hence the robust current limits we put are restricted to this region i.e. above the top black line. The dashed lines show the reach for an identical heating rate experiment that is conducted in a deep mine at 1km depth. Virial DM fractions as small as $f_Q\approx 10^{-12}$ are already ruled using existing heating data for DM masses in the 1-10 GeV mass range. For heavier masses, the terminal velocity is larger and hence the traffic jam densities are smaller. Nonetheless we set limits on DM fractions as small as $f_Q=10^{-3}$ for masses as large as $m_Q\approx 10^5$~GeV.

The limits presented above were derived using data from existing experiments that measure the anomalous heating rate. We next make projections for the future to capture improvements in reducing the heating rate as well as to incorporate mCP detection specific modifications. 

\subsection{Projections}
\begin{figure}
\centering
    \includegraphics[width=\linewidth]{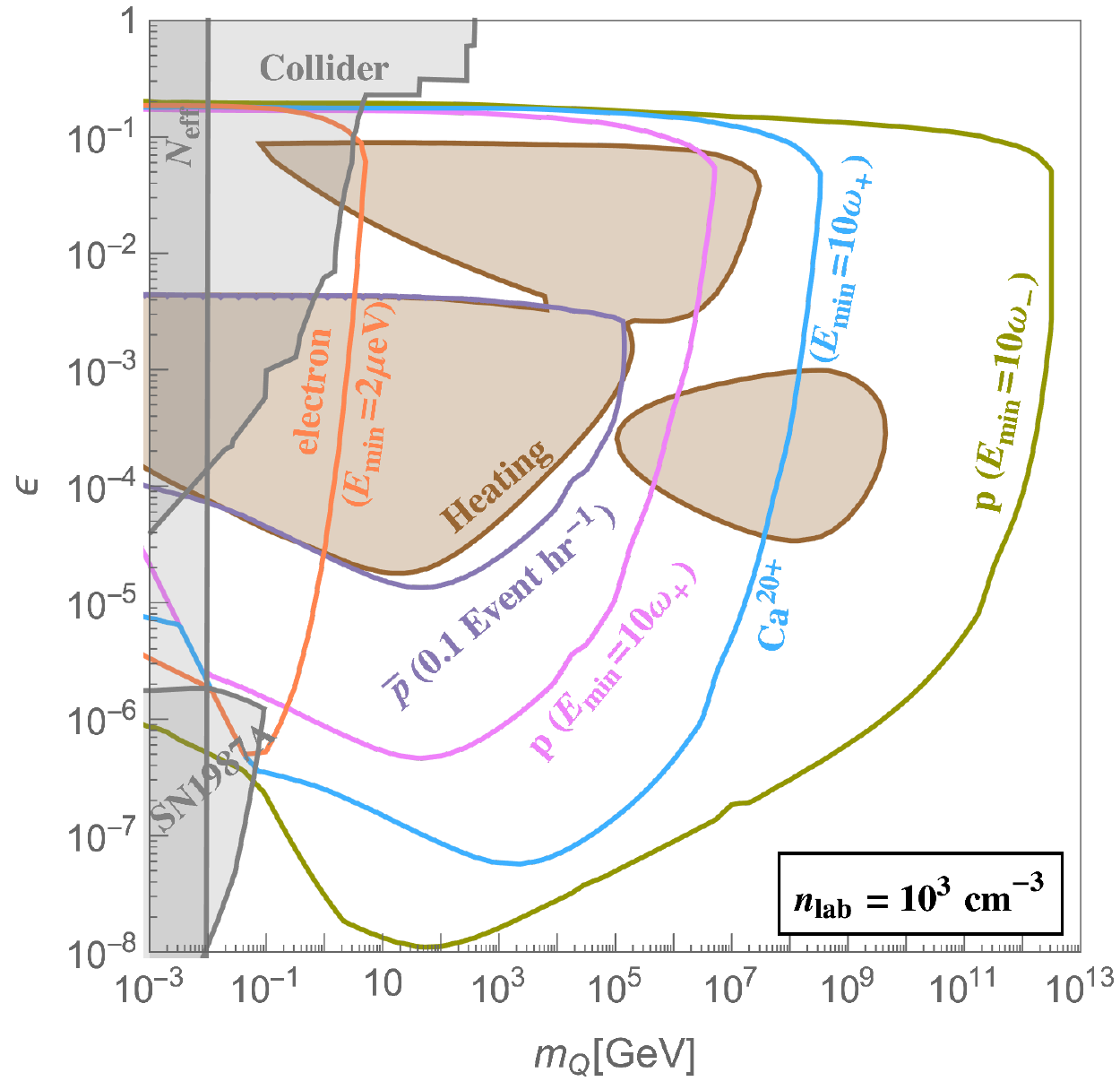} 
    
    \caption{Event Rate limits and projections for $n_{\rm lab}=10^3 \textrm{cm}^{-3}$. Existing limits from heating of the $\omega^-$ mode are shown (brown shaded) from Fig \ref{fig:evsm}.  The projection for a search for single events in the BASE experiment \cite{borchert2019measurement} with energy deposit above $10\omega^+$ and rate 0.1 event/hr are shown (dark blue). Here $\omega_+=77.4$ neV and $\omega_-= 0.050$ neV.  Next, projections are also shown assuming sensitivity to 1 Event/year event rates. In pink, we show sensitivity from the existing setup of Ref.~\cite{borchert2019measurement}. The light blue curve corresponds to swapping a single (anti)-proton with fully ionized Calcium or equivalently trapping 400 Calcium ions in a Coulomb crystal. We also show the reach for a futuristic experiment with energy thresholds of $E_{\rm min}=10\omega_{-}$ in green. Finally, reach from a hypothetical electron trap with trap frequency $\omega=200$~neV and $E_{\rm min}=2~\mu$eV is shown in orange.
    }
    \label{fig:futureproj}
\end{figure}

Limits shown thus far arise from the cumulative heating rate measured. Another promising avenue, is the (non)-observation of individual event rates. In Fig.~\ref{fig:futureproj}, we compare projections from a non-observation of single event with $E_{\rm ion}\ge 10\omega_+$ with the same parameters as existing data in Ref.~\cite{borchert2019measurement} in dark blue with the heating of the much smaller $\omega_-$  in brown for $n_{\rm lab}=10^3~\textrm{cm}^{-3}$. 

We find that this projection is near-identical to the heating limit in brown. Both the heating limits as well as the event rate sensitivity are expected to improve in the future. For e.g. heating rates are known to decrease with larger electrode distance or increasing the frequency of the trapped ion~\cite{brownnutt2015ion}. Whereas the mCP search with the heating rate in the current BASE apparatus is already background limited, the event rate analysis is not. 

It is unclear what the limiting background will be for events with $E_{\min}=10\omega$. The harmonic oscillator selection rules prevent excitation of $E_{\rm ion}>\omega$. Background gas particles at the existing pressure of $3\times 10^{-18}$~mbar and $100\textrm{\AA}^2$ cross-section, correspond to 1 event every 5000 years. The limiting rate will perhaps be electric field noise, whose estimate is unknown. In Fig.~\ref{fig:futureproj} we make projections for various experimental choices for an optimistic choice of $\textrm{1 yr}^{-1}$ event rates for $n_{\rm lab}=10^3 ~\textrm{cm}^{-3}$. In pink, we show projections for a trapped proton keeping the existing energy threshold $E_{\rm min}=10\omega_+$ and other parameters in \cite{borchert2019measurement}. 
Next, we explore the reach for highly charged ions as well as ions in a lattice (see for e.g. Ref.~\cite{micke2020coherent} for recent heating limits from highly charged $^{40}\textrm{Ar}^{13+}$ ions in a lattice). In light blue, we consider the same set up as \cite{borchert2019measurement} but consider the reach for fully ionized Calcium, which enhances the Rutherford scattering cross-section due to large ionic charge. This limit is also  equivalent to 400 ions in a Coulomb crystal observed for 1 year.  In green, we consider the effect of a vast and futuristic improvement in sensitivity to the energy jump of a single event, $E_{\rm min}=10\omega_-$. Finally we consider a hypothetical electron trap, with trap frequency $\omega=0.3$~GHz and $E_{\rm min}$=3~GHz. Despite the large trap frequency required for electron trapping, it is competitive with ions in the small $m_Q$ regime. This is because the momentum transfer is much smaller for electron targets compared to ions for the same energy transfer.

\section{Discussion}
\label{sec:discussion}
Generic cosmologies should produce non-trivial abundances of well motivated stable particles that make up some or all of the observed DM density provided the reheat temperature is high enough. Such a cosmic density of millicharge particles, with large enough charge, can get stopped on Earth and accumulate through the planet's history, forming an overdense, locally thermalized population. In this work, we analyzed the utility of ion traps as detectors of such an mCP population. Ions trapped in harmonic potentials can detect energy deposits as small as neV with low intrinsic backgrounds. We have shown that the existing measurement of heating rates in Penning and Paul traps~\cite{borchert2019measurement,goodwin2016resolved,hite2012100} sets strong limits on a wide range of mCP masses and charges as seen in Fig.\,\ref{fig:evsm}. These limits on the ambient thermalized population improve on existing limits from levitated spheres by several orders of magnitude with no assumptions about binding of mCPs in material. This can be seen in Fig.\,\ref{fig:compexist}.

These limits can in turn be interpreted as limits on the fraction of virial DM existing in mCPs, $f_Q= \frac{\rho_Q}{\rho_{\rm DM}}$. We find in Fig.\,\ref{fig:virial} that fractions as small as $f_Q\gtrsim 10^{-12}$ are ruled out for masses around 1-100 GeV for millicharge around $10^{-3}$. Smaller charges can be probed with a similar setup conducted deep underground. 

Turning to future prospects, while modest improvements in the observed heating rates are expected, greater strides can be made with single event observation as seen in Fig.\,\ref{fig:futureproj}. With event/year sensitivities, a single (anti)proton can improve upon the existing bounds from heating by one order of magnitude with energy thresholds of $\approx 100$ neV. Reducing energy threshold can further increase the parameter space that can be probed.

Another viable direction is using a single fully ionized heavy ion ~\cite{micke2020coherent} which increases the Rutherford cross-section with mCPs. Multiple ions that form a Coulomb crystal also result in increased sensitivity. While we made projections only for the single event rate for a Coulomb crystal, it is reasonable to expect better signal/background discrimination by considering in detail the selection rules for collective excitations of the crystal. Heavier ions will however require improvements in energy resolution in order to be sensitive to the same quantum jump as an (anti)proton experiment. 

Ion experiments have been shown to be transportable~\cite{cao2016transportable,delehaye2018single,gellesch2020transportable} and can thus be conducted at different altitudes; deep underground in mines, at high altitudes or in space will have drastically different mCP induced heating rate due to the large densities at mines and virial densities in space. This can be used for sensitivity to smaller ambient mCP densities as well as a viable background discrimination tool.

Finally, electrons can be stably trapped in deep potential wells. Due to its lower mass, an electron can extract $\approx\frac{m_p}{m_e}$ more energy than a proton for the same momentum transfer. For masses around 1\,GeV and below, electron traps might be a promising alternative to probe low charges. It is important to emphasize the complementarity of different traps. Existing Traps of larger sizes have lower heating rates but feature deeper potentials. Hence, they are suited for probing small charges, whereas microtraps are ideal for larger charges. The choice of the trapped charged SM particle, a heavy ion, a proton or an electron can provide optimal sensitivity to different masses due to kinematic matching. For mCP detection purposes, a large trap with shallow trapping potentials will be optimal. 

Ion trapping has myriad applications including the realization of qubits for a quantum computer. There are significant resources invested in this endeavor which should translate to longer stability, reduced heating rates, scaling to large numbers of ions, as well as the realization of long-term stability in electrons. It is exciting that these improvements translate directly into increased sensitivity of dark matter detection.  
\\
\\
\textbf{Note Added:} While this paper was in preparation, ideas related to detecting mCPs using ion traps appeared in Ref~\cite{Carney:2021irt}. Event rate measurements were explored, which is a subset of the observables discussed in this work. While future projections were made in Ref.~\cite{Carney:2021irt}, our work additionally provides existing limits from the heating rate, which is already measured in many traps. 
While we agree qualitatively with the broad conclusions of the Ref.~\cite{Carney:2021irt} for the future, that ion traps are ideal mCP detectors, we disagree with the quantitative results.
In particular, for the future projections using their parameters, we find different answers for the reach.  The discrepancy may arise from the following differences.
First, in Ref.~\cite{Carney:2021irt} the scattering is treated  as a  Rutherford scatter on a free, stationary ion. Taking into account the initial energy of the ion makes a significant difference. Second, in Ref.~\cite{Carney:2021irt}, the ambient number density is conservatively assumed to be the virial one, whereas, our work incorporates the orders of magnitude increase in local density. Third, we analyze the effect of work functions, trapping potentials and mean-free-paths which are very important at large charge. This  provides a maximum charge above which there is no reach. Fourth, we point out that experiments performed near the surface of the earth have access to only a constrained region in charge vs mass, which can be expanded if the experiment was to instead be performed in a deep mine. 
The initial arxiv version of the publication \cite{Carney:2021irt} takes into account only virialized velocity distributions, so it did not apply to the parameter space we consider where current ion traps have sensitivity.  The subsequent slowdown to room and cryogenic temperatures were incorporated in the published version which just appeared during the final stages of this work, though with the differences noted above.

\acknowledgments
We would like to thank Richard Thompson for useful discussions regarding the particulars of the experimental data in ~\cite{goodwin2016resolved}.  The work of DB and FSK was supported by the Cluster of Excellence ``Precision Physics, Fundamental Interactions, and Structure of Matter'' (PRISMA+ EXC 2118/1) funded by the German Research Foundation (DFG) within the German Excellence Strategy (Project ID 39083149), by the European Research Council (ERC) under the European Union Horizon 2020 research and innovation program (project Dark-OST, grant agreement No 695405), and by the DFG Reinhart Koselleck project. SU acknowleges support by RIKEN and the Max Planck, RIKEN, PTB Center for Time, Constants and Fundamental Symmetries. 
CS acknowledges support by the ERC (project STEP, grant agreement No 852818) and the Institute of Physics in Mainz. 
PWG and HR acknowledge support from the Simons Investigator Award 824870, DOE Grant DE-SC0012012, NSF Grant PHY-2014215, DOE HEP QuantISED award \#100495, and the Gordon and Betty Moore Foundation Grant GBMF7946.  This work was also supported by the U.S.~Department of Energy, Office of Science, National Quantum Information (NQI) Science Research Centers through the Fermilab SQMS NQI Center.

\onecolumngrid

\appendix
\section{Passage through trap}
 \label{Appendix:passage}
 In this appendix we provide the details of the passage through the trap, as is relevant to the discussion in Sec.~\ref{trappassage}. The aim is to relate the ambient density in the lab $n_{\rm lab}$ to the density at the point where the ion is trapped, $n^Q_{\rm ion}$.
 \subsection{Number density of mCPs in trap}
 \label{Appendix: number density}
 
 Common to both Penning and Paul traps is the metallic high vacuum container that sets the wall temperature $T_{\rm wall}$. As described earlier, the Penning and Paul traps differ in the mechanism to confine ions in the perpendicular plane with the magnetic and RF fields preventing mCP propagation in the perpendicular direction for large enough $\epsilon$. On the other hand, for propagation along the axial direction, one can ignore these fields. However, in the axial direction, some other important effects need to be considered. In the order they occur they are,  thermalizing in walls, penetration of the double wall, effect of vacuum pump, static electric potential. 
 In order to capture these effects, let us consider the following path for the mCPs:
\begin{enumerate}
    \item Outside at temperature $T_{\rm room}$ with mCP density $n_{\rm lab}$
    \item Metal A with barrier=$\epsilon \phi_1$ at temperature $T_{\rm wall}$ with density $n_A$ and volume $V_{\rm A}$
    \item buffer vacuum with effective temperature $T_{\rm wall}$ with density $n_{\rm buf}$ and volume $V_{\rm buf}$
    \item Metal B with barrier=$\epsilon \phi_2$ at temperature $T_{\rm wall}$ with density $n_{B}$ and volume $V_{\rm B}$
    \item vacuum inside with density $n_{\rm vac}$ and has volume $V_{\rm vac}$
\end{enumerate}
In practice, there could be a series of metals at decreasing temperatures and varied work functions as described in Section.~\ref{sec:traps}. We find that the smallest density inside the trap is obtained by two adjacent metals, both cold and decreasing work functions, and hence work with this simplified model, which nonetheless captures the inherent suppression. 
Note that the barrier for a positively charged particle to enter the metal from outside $\epsilon \phi$ is described in detail in Appendix.~\ref{Appendix:work function}.

The change in local number density in any segment is given by the sum of the fluxes along each wall in the axial direction, ignoring the perpendicular direction. For the metal A, the net incoming flux is,  
\begin{equation}
\textrm{flux}_{\rm in}^A=\left(n_{\rm lab} e^{-\frac{\epsilon \phi_1}{T_{\rm room}}}\sqrt{\frac{T_{\rm room}}{2\pi m_Q}} +n_{\rm buf} e^{-\frac{\epsilon \phi_1}{T_{\rm wall}}}\sqrt{\frac{T_{\rm wall}}{2\pi m_Q}} \right).
\end{equation}
Here, the $\sqrt{\frac{T_{\rm room}}{2\pi m_Q}}$ comes from the average velocity and the exponent from needing to penetrate the double layer. We can similarly write the outgoing flux, which gives,
\begin{equation}
\textrm{flux}_{\rm out}^A=2 n_A \sqrt{\frac{T_{\rm wall}}{2\pi m_Q}}.
\end{equation}
Here, the exponent is absent because the double layer pushes out all the positive charges that reach the boundary and hence the flux is independent of the presence of the double layer. We can then write $\dot{n}=\frac{A_{\rm in}}{V}\textrm{flux}_{\rm in}-\frac{A_{\rm out}}{V}\textrm{flux}_{\rm out}$ for the different parts:

\begin{align}
   \dot{n}_A&=\frac{A_{\rm wall}}{V_A}\left(n_{\rm lab} e^{-\frac{\epsilon \phi_1}{T_{\rm room}}}\sqrt{\frac{T_{\rm room}}{2\pi m_Q}} +n_{\rm buf} e^{-\frac{\epsilon \phi_1}{T_{\rm wall}}}\sqrt{\frac{T_{\rm wall}}{2\pi m_Q}} -2 n_A \sqrt{\frac{T_{\rm wall}}{2\pi m_Q}}\right), \nonumber \\
    \dot{n}_{\rm buf} &=\frac{A_{\rm wall}}{ V_{\rm buf}}\sqrt{\frac{T_{\rm wall}}{2\pi m_Q}}\left(n_A+n_B -n_{\rm buf}  \left(e^{-\frac{\epsilon \phi_1}{T_{\rm wall}}}+e^{-\frac{\epsilon \phi_2}{T_{\rm wall}}}\right)\right), 
    \nonumber \\
    \dot{n}_{B}&= \frac{A_{\rm wall}}{V_{\rm ion}}\sqrt{\frac{T_{\rm wall}}{2\pi m_Q}}\left(\left(n_{\rm buf}+n_{\rm vac}\right)e^{-\frac{\epsilon \phi_2}{T_{\rm wall}}}-2n_{B}\right),\nonumber \\
   \dot{n}_{\rm vac}&=\sqrt{\frac{T_{\rm wall}}{2\pi m_Q}}\left(A_{\rm wall} n_B  -n_{\rm vac}\left(A_{\rm wall}e^{-\frac{\epsilon \phi_2}{T_{\rm wall}}}+A_{\rm fan}\right)\right)/V_{\rm vac}.
   \label{eq:bigeq}
\end{align}
In the last line, the effect of the vacuum pump is taken into account by putting in an area $A_{\rm fan}$ for the vacuum fan. The assumption here is that all mCPs that hit the area, $A_{\rm fan}$ are effectively evacuated from the vacuum region. 
The steady state solution to this is,
\begin{align}
 n_{\rm vac}^{\rm eqbm}=n_\text{lab}\frac{\sqrt{T_{\rm room}}}{\sqrt{T_{\rm wall}}}\frac{ e^{-\frac{\epsilon \phi_1}{T_{\rm room}}+\frac{\epsilon \phi_1}{T_{\rm wall}}}}{ \left(2 \frac{A_\text{fan}}{A_{\rm wall}}e^{\frac{
 \epsilon  \left(\phi_1+\phi_2\right)}{T_{\rm wall}}}+ 1\right)} \equiv n_{\rm on} \quad \quad &&& \textrm{Fan on}
\end{align}
If the fan is turned off, the equilibrium density is set by $A_{\rm fan}\rightarrow 0$, and given by,
\begin{equation}
  n_{\rm off}=  n_\text{lab}\frac{\sqrt{T_{\rm room}}}{\sqrt{T_{\rm wall}}}\left( e^{\frac{\epsilon \phi_1}{T_{\rm wall}}-\frac{\epsilon \phi_1}{T_{\rm room}}}\right) \quad \quad \quad \quad \quad \quad \quad  \quad \quad \quad  \quad \quad \quad \quad \quad \quad  \textrm{Fan off}
\end{equation}

This contains an exponential enhancement in the limit $T_{\rm wall} \ll T_{\rm room}$. This happens because mCPs readily leak from the metal into the vacuum, subsequently get cooled to $T_{\rm wall}$ and now have a barrier to re-enter the metal. This exponential enhancement happens only with a hermetic metal container and is a potentially promising mechanism to greatly enhance the number density inside the trap. Since the hermeticity of the containers are unknown and in the spirit of being conservative, we never allow densities to put constraints to go above $n_{\rm lab} \sqrt{\frac{T_{\rm room}}{T_{\rm wall}}}$.

Let us now discuss the dynamics after the fan is turned off. $n_{\rm vac}$ is somewhere in between $n_{\rm off}$ and $n_{\rm on}$  The mCPs in metal B, diffuse into vacuum after the fan is turned off. Metal B has number density relatively independent of on/off. Solving Eq.\,\ref{eq:bigeq}, it is given by,
\begin{align}
n_B=\frac{\sqrt{T_{\rm room}}}{\sqrt{T_{\rm wall}}}\frac{n_{\rm lab}}{2}e^{-\frac{\epsilon \phi_1}{T_{\rm room}}} e^{-\frac{\epsilon \left(\phi_2-\phi_1\right)}{T_{\rm wall}}}.
\end{align}
The second term is an exponential enhancement.
Assuming the fan is turned off at time=0, the number density in the vacuum after time $\tau$ is,
\begin{align}
\label{eqn:filling trap density}
  n_{\rm vac}(\tau)=n_B \frac{\tau}{L_{\rm vac}}\sqrt{\frac{T_{\rm wall}}{2\pi m_Q} }+n_{\rm on},   
\end{align}
where $L_{\rm vac}=\frac{V_{\rm vac}}{A_{\rm vac}}$.
This was solved in the small $\tau$ limit, the maximum it can reach is $n_{\rm off}$. As noted earlier  $n_{\rm off}$ contains an exponential enhancement which we cut off at $n_{\rm lab} \sqrt{\frac{T_{\rm room}}{T_{\rm wall}}}$. Finally then, the ambient number density in the trap vacuum is, 
\begin{align}
\label{eqn:final trap density}
    n_{\rm trap}=\textrm{Min}\left[n_{\rm vac}(\tau), n_{\rm off}, n_{\rm lab} \sqrt{\frac{T_{\rm room}}{T_{\rm wall}}}\right],
\end{align}
where  
$n_{\rm vac}(\tau)$ is the number density as the trap is filling with mCPs before it reaches equilibrium as calculated in Eq.\,\eqref{eqn:filling trap density}.

The mCPs are thermalized to $T_{\rm wall}$ inside the trap.  Then to find the number density at the position of the ion we must take into account  the axial potential barrier $V_z$.  To do this we take a Boltzmann factor on top of the average number density $n_{\rm trap}$ inside the trap to finally find the number density at the position of the ion to be
\begin{equation}
n^Q_{\rm ion}  = e^{- \frac{\epsilon V_z}{T_{\rm wall}}} \, n_{\rm trap}.
 \label{eqn:qion}
\end{equation}
Note that values of $V_z$ are listed in Tab.\,\ref{table:data}. $n^Q_{\rm ion}$ is the density we use in the rate calculations.

\subsection{Work function of metals}
\label{Appendix:work function}

Here we consider the effect of the work function of a metal on the passage of mCPs through that metal.  This is only relevant for mCPs of relatively large charge ($\epsilon \gtrsim 10^{-2}$) but for those it can be a large effect.  The work function for mCPs, which we will call $\epsilon \phi$, is not simply $\epsilon$ times the work function for electrons.  For electrons the work function arises from several contributions (of varying sign) including e.g.~the binding to the lattice of nuclei, the Fermi sea of other electrons, and surface effects such as the ``double layer" or the image charge potential.  Several of these do not apply to mCPs and so the final answer for the work function of an mCP in a metal is significantly different from $\epsilon$ times the work function for electrons.

The density of mCPs is low enough that any Fermi sea of other mCPs (if they are fermions) is not relevant.  For electrons removing the Fermi sea would increase the work function, making them more deeply bound to the metal than their normal work function.

The image charge of the mCP also has charge $\epsilon e$ and so the potential energy between mCP and image charge is $\propto \epsilon^2$ and is irrelevant for $\epsilon \lesssim 0.1$ which is the region we consider.  For electrons, the image charge force is attractive and so removing it makes electrons less deeply bound compared to their normal work function.

Further we are considering positive mCPs so they do not bind to nuclei.  Of-course they will have an electromagnetic interaction with the positively charged (point-like) nuclei and the negatively charged (relatively uniform) sea of electrons.  The only negative potential energy contribution could come from binding with electrons, but such a bound state would be much larger than the normal size of an atom (or interatomic spacing) by a factor $\epsilon^{-1}$.  Within such a large distance there will be many nuclei and electrons and thus a roughly zero net charge density.  Note that the mCP itself is significantly heavier and thus of smaller wavelength than an electron in the entire parameter space we consider.  So its wavefunction is quite different than the electron's wavefunction.  The positive mCP is mainly repelled by the repulsive potentials of positively charged nuclei.  So at most we would expect that the effect of interaction would be a repulsion of the mCP, namely a positive potential energy relative to infinity.  But this is unlikely to be very significant.

The `double layer' surface effect is relevant for mCPs. Since the metal has negative charge density extending outside the region which contains the positive nuclei, it acts a bit like a capacitor around the edges of the metal.  Namely it has a potential energy barrier for the mCPs which is simply $\epsilon$ times the magnitude of the usual double layer contribution to the work function for electrons.  We will thus use this as our estimate for the work function (potential energy barrier) for mCPs.

In Tab.\,\ref{tab:work functions} we list the usual electron work functions (WF) for several relevant metals surrounding the experiments we consider.  We also list the double layer contribution (DL) and the remainder of the work function ($\mu$). Wherever available this data is taken directly from~\cite{lang1971theory} and the numbers for the rest of the metals are interpolated from the data available in~{\cite{lang1971theory} for different $r_s$.  Note that we have taken conventions in which:
\begin{equation}
\text{WF} = \text{DL} - \mu.
\end{equation}
We then take the potential energy barrier for the mCPs crossing a metal to be
\begin{equation}
\epsilon \phi = \epsilon \times \text{DL},
\end{equation}
where we have taken conventions where DL is positive.

\begin{table}[ht]
\centering
\begin{tabular}{ ||c|c|c|c|c|| }
\hline
 Element & $r_s$ & $\mu$ [eV] & DL [eV] & WF [eV] \\ 
 \hline \hline
 Copper & 2.67 & -0.45 & 3.19 & 3.65\\
 Steel & 2.12 & 2.05 & 5.91& 3.86 \\
 Aluminium & 2.07 &2.39 &6.26 & 3.87\\
 Gold &3.01 &-1.20 &2.30 &3.5\\
 Nickel & 2.59&-0.20 &3.47 &3.68 \\
 \hline
\end{tabular}
\caption{  The electron work functions (WF) for several relevant metals surrounding the experiments we consider. If not available in~\cite{lang1971theory}, interpolation is used to estimate the numbers using the known $r_s$ values. The double layer contribution (DL) and the remainder of the work function ($\mu$) are also shown.}
\label{tab:work functions}
\end{table}

As noted above in Appendix \ref{Appendix: number density} the relevant effect of the work function of the metals around the experiment comes from the case where the experiment is encased in two different metals and the  work function for mCPs of the inner metal is larger (in magnitude) then that of the outer one.
Note that for many of the experiments, the metals enclosing them are often grounded.  This slightly modifies the above difference in work functions.
If the two metals were not electrically connected then we would just have the difference in the two mCP work functions $\Delta \phi = \phi_1 - \phi_2 = \text{DL}_1 - \text{DL}_2$.
But if they are connected by a wire (or both grounded) then you also subtract from this difference in double layers the difference in electron work functions (WF), thus ending up with just the difference of the mu's.  This is because of the contact potential effect between two metals.  Thus in this case the potential barrier height difference between the two metals is
\begin{equation}
\Delta U = \epsilon \left( \left( \text{DL}_1 - \text{DL}_2 \right) - \left( \text{WF}_1 - \text{WF}_2 \right) \right) = \epsilon \left( \mu_1 - \mu_2 \right). 
\end{equation}
In this case, this would actually be what enters all the equations in Appendix \ref{Appendix: number density} instead of the quantity $\epsilon \left( \phi_1 - \phi_2 \right)$.  However since the work functions of the relevant metals are not very different, this does not appreciably change our answer.

There are many paths that an MCP may take to enter each actual experiment with multiple different metals to pass through.  Further, while the overall work functions of each metal are measured quantities, the part of the work function which comes from the double layer must be calculated and may have some uncertainty to it.  Thus in order to avoid all this complication we simply take very conservative estimates for the mCP work functions.  From Tab.\,\ref{tab:work functions} it can be seen that the differences in work functions that are relevant for any of the experiments is at most $\sim 3 \, \eV$.  Therefore we assume $\phi_2 - \phi_1 = 3 \, \eV$ for all equations in Appendix \ref{Appendix: number density}.  This will set the top of the excluded regions in Fig.~\ref{fig:evsm} and the right side in Fig.~\ref{fig:compexist}.  Further we assume $\phi_1 = 3 \, \eV$ (and so $\phi_2 = 6 \, \eV$) wherever relevant, which is conservative.  
\section{Heating Rates}
\label{app1}
Consider mCPs $m_Q$ scattering with an ion target. Let us assume that their velocities are $\mathbf{v_Q}$ and $\mathbf{v_{\rm ion}}$. The CM velocity of each particle is given by
\begin{equation}
    \mathbf{v_{\rm ion}}^{\rm CM,i}=(\mathbf{v_{\rm ion}}-\mathbf{v_Q})\frac{m_{Q}}{m_{\rm ion}+m_Q},
\end{equation}
and the final velocity is
\begin{equation}
    \mathbf{v_{\rm ion}}^{\rm CM,f}=v_{\rm ion}^{\rm CM,i} \hat{n}. 
\end{equation}
Here, we use the fact that the magnitude of the velocity does not change in the CM frame. The final velocity of the ion is assumed to be in the $\mathbf{\hat{n}}$ direction. 

The change in velocity in any frame is,
\begin{align}
    \mathbf{\Delta v_{\rm ion}}&=\frac{m_Q}{m_{\rm ion}+m_Q} |\mathbf{v_{\rm ion}}-\mathbf{v_Q}| \left(\mathbf{\hat{n}}-\frac{\mathbf{v_{\rm ion}}-\mathbf{v_Q}}{|v_{\rm ion}-v_Q|}\right) \nonumber \\
    &=\frac{m_Q}{m_{\rm ion}+m_Q}\left[ \left(\cos \theta-1\right) \left(\mathbf{v_{\rm ion}}-\mathbf{v_Q}\right)+\sin \theta |\mathbf{v_{\rm ion}}-\mathbf{v_Q}| \mathbf{n_{\perp}}\right]. 
\end{align}
We can drop the $\sin \theta$ piece because it averages to zero. 
The CM velocity $\mathbf{v_{CM}}$ is given by,
\begin{equation}
    \mathbf{v_{CM}}=\frac{\left(m_{\rm ion} \mathbf{v_{\rm ion}}+m_Q \mathbf{v_Q}\right)}{m_{\rm ion}+m_Q}.
\end{equation}
The energy transfer is given by,
\begin{equation}
    \Delta E_{\rm ion} = m_{\rm ion} \mathbf{v_{CM}}.\mathbf{\Delta v_{\rm ion}}.
\end{equation}

We simplify this equation in the simple case of free ion targets, in order to learn qualitative features. In the free ion limit, there is no minimum energy transfer $E_{\rm min}$. Hence, the energy transfer rate is given by,

\begin{align}
    \dot{H}&=n_Q \langle \sigma \Delta E_{\rm ion} v_{\rm rel} \rangle \nonumber \\
    &=\frac{n_Q m_Q m_{\rm ion}}{m_{\rm ion}+m_Q} \int d^3 v_Q g_Q \int d^3 v_{\rm ion} g_{\rm ion}\int d \cos \theta \frac{d\sigma}{d\cos \theta} \left(1-\cos \theta\right)|\mathbf{v_Q}-\mathbf{v_{\rm ion}}| \mathbf{v_{ CM}}.\left(\mathbf{v_Q}-\mathbf{v_{\rm ion}}\right) \nonumber \\
    &=\frac{n_Q m_Q m_{\rm ion}}{m_{\rm ion}+m_Q} \int d^3 v_Q g_Q \int d^3 v_{\rm ion} g_{\rm ion} |\mathbf{v_Q}-\mathbf{v_{\rm ion}}|  \mathbf{v_{ CM}}.\left(\mathbf{v_Q}-\mathbf{v_{\rm ion}}\right) \sigma_{\rm ion}\left(|\mathbf{v_Q}-\mathbf{v_{\rm ion}}|\right). 
\end{align}
Here $\sigma_{\rm ion}$ is the transfer cross-section for Rutherford Scattering. 

We now do a change of variable,
\begin{align}
\mathbf{v_{rel}}&=\mathbf{v_Q}-\mathbf{v_{\rm ion}} \nonumber \\
\mathbf{v_m}&=\frac{\frac{m_Q}{T_Q} \mathbf{v_Q}+\frac{m_{\rm ion}}{T_{\rm ion}} \mathbf{v_{\rm ion}}}{\frac{m_Q}{T_Q}+\frac{m_{\rm ion}}{T_{\rm ion}}}. 
\end{align}

Then,
\begin{equation}
    \int d^3 v_Q g_Q \int d^3 v_{\rm ion} g_{\rm ion}= \int d^3 v_{\rm rel} f_{\rm rel} \int d^3 v_m f_m. 
\end{equation}
The thermal width of $f_{\rm rel}$ is given by $\frac{T_Q}{m_Q}+\frac{T_{\rm ion}}{m_{\rm ion}}$.
Also,
\begin{align}
    \mathbf{v_{CM}}&= \mathbf{v_m} +\frac{m_{\rm ion} m_Q\left(T_Q-T_{\rm ion}\right) }{\left(m_{\rm ion}+m_Q\right)\left(m_Q T_{\rm ion}+m_{\rm ion} T_Q\right)}\mathbf{v_{rel}}. 
\end{align}
Then,
\begin{align}
    \dot{H}=\frac{n_Q m_Q m_{\rm ion}}{m_{\rm ion}+m_Q} \int d^3 v_{\rm rel} f_{\rm rel} \int d^3 v_m f_m |\mathbf{v_{\rm rel}}|  (\mathbf{v_m} +\frac{m_{\rm ion} m_Q\left(T_Q-T_{\rm ion}\right) }{\left(m_{\rm ion}+m_Q\right)\left(m_Q T_{\rm ion}+m_{\rm ion} T_Q\right)}\mathbf{v_{rel}}).\mathbf{v_{rel}} \sigma_{\rm ion}\left(|\mathbf{v_{rel}}|\right). 
\end{align}
Now the term proportional to just $\mathbf{v_m}$ is odd and hence its integral is zero, the rest is independent of $v_m$ and hence the Boltzmann integral $v_m$ gives 1. This gives,
\begin{equation}
    \dot{H}=\frac{n_Q m_{\rm ion}^2 m_Q^2\left(T_Q-T_{\rm ion}\right)}{\left(m_{\rm ion}+m_Q\right)^2\left(m_Q T_{\rm ion}+m_{\rm ion} T_Q\right)} \int d^3 v_{\rm rel} f_{\rm rel} v_{\rm rel}^3 \sigma_t(v_{\rm rel}). 
\end{equation}
When $\sigma_t=\frac{\sigma_0}{v_{\rm rel}^4}$, then,
\begin{equation}
    \int d^3 v_{\rm rel} \frac{f_{\rm rel}}{v_{\rm rel}} = \left(\frac{2}{\pi}\frac{m_{\rm ion} m_Q}{(m_Q T_{\rm ion}+m_{\rm ion} T_Q)}\right)^\frac{1}{2}.
\end{equation}
Defining $\frac{T_{\rm ion}}{m_{\rm ion}}+\frac{T_Q}{m_Q}=u_{\rm th}^2$, we get,
\begin{equation}
    \dot{H}=\sqrt{\frac{2}{\pi}}\frac{n_Q m_Q m_{\rm ion} (T_Q-T_{\rm ion})}{(m_{\rm ion}+m_Q)^2}\frac{\sigma_0}{u_{\rm th}^3}.
\end{equation}
We see that the heating rate is proportional to the difference in temperatures $T_Q-T_{\rm ion}$. Furthermore the rate is also proportional to $u_{\rm th}^{-3}$.

However, the ions are in a harmonic oscillator potential and hence there is a minimum energy $E_{\rm min}$ such that
\begin{align}
    \Delta E_{\rm ion}>E_{\rm min}.
\end{align}

In this case, the heating rate is instead, 

\begin{align}
    \dot{H}&=n^Q_{\rm ion}  \int d^3 \mathbf{v_Q} g_Q  \int d^3 \mathbf{v_{\rm ion}} g_{\rm ion}  \int d \Omega |v_Q-v_{\rm ion}|  \nonumber \\ &\frac{d\sigma} {d \Omega}E_{\rm ion} \Theta \left(|E_{\rm ion}|-E_{\rm thr}\right) \Theta\left(E_{\rm samp}-|E_{\rm ion}|\right). 
\end{align}

\bibliography{biblio}

\begin{thebibliography}{61}%
\makeatletter
\providecommand \@ifxundefined [1]{%
 \@ifx{#1\undefined}
}%
\providecommand \@ifnum [1]{%
 \ifnum #1\expandafter \@firstoftwo
 \else \expandafter \@secondoftwo
 \fi
}%
\providecommand \@ifx [1]{%
 \ifx #1\expandafter \@firstoftwo
 \else \expandafter \@secondoftwo
 \fi
}%
\providecommand \natexlab [1]{#1}%
\providecommand \enquote  [1]{``#1''}%
\providecommand \bibnamefont  [1]{#1}%
\providecommand \bibfnamefont [1]{#1}%
\providecommand \citenamefont [1]{#1}%
\providecommand \href@noop [0]{\@secondoftwo}%
\providecommand \href [0]{\begingroup \@sanitize@url \@href}%
\providecommand \@href[1]{\@@startlink{#1}\@@href}%
\providecommand \@@href[1]{\endgroup#1\@@endlink}%
\providecommand \@sanitize@url [0]{\catcode `\\12\catcode `\$12\catcode
  `\&12\catcode `\#12\catcode `\^12\catcode `\_12\catcode `\%12\relax}%
\providecommand \@@startlink[1]{}%
\providecommand \@@endlink[0]{}%
\providecommand \url  [0]{\begingroup\@sanitize@url \@url }%
\providecommand \@url [1]{\endgroup\@href {#1}{\urlprefix }}%
\providecommand \urlprefix  [0]{URL }%
\providecommand \Eprint [0]{\href }%
\providecommand \doibase [0]{http://dx.doi.org/}%
\providecommand \selectlanguage [0]{\@gobble}%
\providecommand \bibinfo  [0]{\@secondoftwo}%
\providecommand \bibfield  [0]{\@secondoftwo}%
\providecommand \translation [1]{[#1]}%
\providecommand \BibitemOpen [0]{}%
\providecommand \bibitemStop [0]{}%
\providecommand \bibitemNoStop [0]{.\EOS\space}%
\providecommand \EOS [0]{\spacefactor3000\relax}%
\providecommand \BibitemShut  [1]{\csname bibitem#1\endcsname}%
\let\auto@bib@innerbib\@empty
\bibitem [{\citenamefont {Izaguirre}\ and\ \citenamefont
  {Yavin}(2015)}]{Izaguirre:2015eya}%
  \BibitemOpen
  \bibfield  {author} {\bibinfo {author} {\bibfnamefont {E.}~\bibnamefont
  {Izaguirre}}\ and\ \bibinfo {author} {\bibfnamefont {I.}~\bibnamefont
  {Yavin}},\ }\href {\doibase 10.1103/PhysRevD.92.035014} {\bibfield  {journal}
  {\bibinfo  {journal} {Phys. Rev. D}\ }\textbf {\bibinfo {volume} {92}},\
  \bibinfo {pages} {035014} (\bibinfo {year} {2015})},\ \Eprint
  {http://arxiv.org/abs/1506.04760} {arXiv:1506.04760 [hep-ph]} \BibitemShut
  {NoStop}%
\bibitem [{\citenamefont {Akers}\ \emph {et~al.}(1995)\citenamefont {Akers},
  \citenamefont {Alexander}, \citenamefont {Allison}, \citenamefont {Ametewee},
  \citenamefont {Anderson}, \citenamefont {Arcelli}, \citenamefont {Asai},
  \citenamefont {Axen}, \citenamefont {Azuelos}, \citenamefont {Ball} \emph
  {et~al.}}]{akers1995search}%
  \BibitemOpen
  \bibfield  {author} {\bibinfo {author} {\bibfnamefont {R.}~\bibnamefont
  {Akers}}, \bibinfo {author} {\bibfnamefont {G.}~\bibnamefont {Alexander}},
  \bibinfo {author} {\bibfnamefont {J.}~\bibnamefont {Allison}}, \bibinfo
  {author} {\bibfnamefont {K.}~\bibnamefont {Ametewee}}, \bibinfo {author}
  {\bibfnamefont {K.}~\bibnamefont {Anderson}}, \bibinfo {author}
  {\bibfnamefont {S.}~\bibnamefont {Arcelli}}, \bibinfo {author} {\bibfnamefont
  {S.}~\bibnamefont {Asai}}, \bibinfo {author} {\bibfnamefont {D.}~\bibnamefont
  {Axen}}, \bibinfo {author} {\bibfnamefont {G.}~\bibnamefont {Azuelos}},
  \bibinfo {author} {\bibfnamefont {A.}~\bibnamefont {Ball}},  \emph {et~al.},\
  }\href@noop {} {\bibfield  {journal} {\bibinfo  {journal} {Zeitschrift
  f{\"u}r Physik C Particles and Fields}\ }\textbf {\bibinfo {volume} {67}},\
  \bibinfo {pages} {203} (\bibinfo {year} {1995})}\BibitemShut {NoStop}%
\bibitem [{\citenamefont {Prinz}\ \emph {et~al.}(1998)\citenamefont {Prinz}
  \emph {et~al.}}]{Prinz:1998ua}%
  \BibitemOpen
  \bibfield  {author} {\bibinfo {author} {\bibfnamefont {A.~A.}\ \bibnamefont
  {Prinz}} \emph {et~al.},\ }\href {\doibase 10.1103/PhysRevLett.81.1175}
  {\bibfield  {journal} {\bibinfo  {journal} {Phys. Rev. Lett.}\ }\textbf
  {\bibinfo {volume} {81}},\ \bibinfo {pages} {1175} (\bibinfo {year}
  {1998})},\ \Eprint {http://arxiv.org/abs/hep-ex/9804008}
  {arXiv:hep-ex/9804008} \BibitemShut {NoStop}%
\bibitem [{\citenamefont {Magill}\ \emph {et~al.}(2019)\citenamefont {Magill},
  \citenamefont {Plestid}, \citenamefont {Pospelov},\ and\ \citenamefont
  {Tsai}}]{Magill:2018tbb}%
  \BibitemOpen
  \bibfield  {author} {\bibinfo {author} {\bibfnamefont {G.}~\bibnamefont
  {Magill}}, \bibinfo {author} {\bibfnamefont {R.}~\bibnamefont {Plestid}},
  \bibinfo {author} {\bibfnamefont {M.}~\bibnamefont {Pospelov}}, \ and\
  \bibinfo {author} {\bibfnamefont {Y.-D.}\ \bibnamefont {Tsai}},\ }\href
  {\doibase 10.1103/PhysRevLett.122.071801} {\bibfield  {journal} {\bibinfo
  {journal} {Phys. Rev. Lett.}\ }\textbf {\bibinfo {volume} {122}},\ \bibinfo
  {pages} {071801} (\bibinfo {year} {2019})},\ \Eprint
  {http://arxiv.org/abs/1806.03310} {arXiv:1806.03310 [hep-ph]} \BibitemShut
  {NoStop}%
\bibitem [{\citenamefont {Acciarri}\ \emph {et~al.}(2020)\citenamefont
  {Acciarri} \emph {et~al.}}]{Acciarri:2019jly}%
  \BibitemOpen
  \bibfield  {author} {\bibinfo {author} {\bibfnamefont {R.}~\bibnamefont
  {Acciarri}} \emph {et~al.} (\bibinfo {collaboration} {ArgoNeuT}),\ }\href
  {\doibase 10.1103/PhysRevLett.124.131801} {\bibfield  {journal} {\bibinfo
  {journal} {Phys. Rev. Lett.}\ }\textbf {\bibinfo {volume} {124}},\ \bibinfo
  {pages} {131801} (\bibinfo {year} {2020})},\ \Eprint
  {http://arxiv.org/abs/1911.07996} {arXiv:1911.07996 [hep-ex]} \BibitemShut
  {NoStop}%
\bibitem [{\citenamefont {Ball}\ \emph {et~al.}(2020)\citenamefont {Ball} \emph
  {et~al.}}]{Ball:2020dnx}%
  \BibitemOpen
  \bibfield  {author} {\bibinfo {author} {\bibfnamefont {A.}~\bibnamefont
  {Ball}} \emph {et~al.},\ }\href {\doibase 10.1103/PhysRevD.102.032002}
  {\bibfield  {journal} {\bibinfo  {journal} {Phys. Rev. D}\ }\textbf {\bibinfo
  {volume} {102}},\ \bibinfo {pages} {032002} (\bibinfo {year} {2020})},\
  \Eprint {http://arxiv.org/abs/2005.06518} {arXiv:2005.06518 [hep-ex]}
  \BibitemShut {NoStop}%
\bibitem [{\citenamefont {Marocco}\ and\ \citenamefont
  {Sarkar}(2021)}]{Marocco:2020dqu}%
  \BibitemOpen
  \bibfield  {author} {\bibinfo {author} {\bibfnamefont {G.}~\bibnamefont
  {Marocco}}\ and\ \bibinfo {author} {\bibfnamefont {S.}~\bibnamefont
  {Sarkar}},\ }\href {\doibase 10.21468/SciPostPhys.10.2.043} {\bibfield
  {journal} {\bibinfo  {journal} {SciPost Phys.}\ }\textbf {\bibinfo {volume}
  {10}},\ \bibinfo {pages} {043} (\bibinfo {year} {2021})},\ \Eprint
  {http://arxiv.org/abs/2011.08153} {arXiv:2011.08153 [hep-ph]} \BibitemShut
  {NoStop}%
\bibitem [{\citenamefont {Ball}\ \emph {et~al.}(2016)\citenamefont {Ball} \emph
  {et~al.}}]{Ball:2016zrp}%
  \BibitemOpen
  \bibfield  {author} {\bibinfo {author} {\bibfnamefont {A.}~\bibnamefont
  {Ball}} \emph {et~al.},\ }\href@noop {} {\  (\bibinfo {year} {2016})},\
  \Eprint {http://arxiv.org/abs/1607.04669} {arXiv:1607.04669
  [physics.ins-det]} \BibitemShut {NoStop}%
\bibitem [{\citenamefont {Berlin}\ \emph {et~al.}(2019)\citenamefont {Berlin},
  \citenamefont {Blinov}, \citenamefont {Krnjaic}, \citenamefont {Schuster},\
  and\ \citenamefont {Toro}}]{Berlin:2018bsc}%
  \BibitemOpen
  \bibfield  {author} {\bibinfo {author} {\bibfnamefont {A.}~\bibnamefont
  {Berlin}}, \bibinfo {author} {\bibfnamefont {N.}~\bibnamefont {Blinov}},
  \bibinfo {author} {\bibfnamefont {G.}~\bibnamefont {Krnjaic}}, \bibinfo
  {author} {\bibfnamefont {P.}~\bibnamefont {Schuster}}, \ and\ \bibinfo
  {author} {\bibfnamefont {N.}~\bibnamefont {Toro}},\ }\href {\doibase
  10.1103/PhysRevD.99.075001} {\bibfield  {journal} {\bibinfo  {journal} {Phys.
  Rev. D}\ }\textbf {\bibinfo {volume} {99}},\ \bibinfo {pages} {075001}
  (\bibinfo {year} {2019})},\ \Eprint {http://arxiv.org/abs/1807.01730}
  {arXiv:1807.01730 [hep-ph]} \BibitemShut {NoStop}%
\bibitem [{\citenamefont {Kelly}\ and\ \citenamefont
  {Tsai}(2019)}]{Kelly:2018brz}%
  \BibitemOpen
  \bibfield  {author} {\bibinfo {author} {\bibfnamefont {K.~J.}\ \bibnamefont
  {Kelly}}\ and\ \bibinfo {author} {\bibfnamefont {Y.-D.}\ \bibnamefont
  {Tsai}},\ }\href {\doibase 10.1103/PhysRevD.100.015043} {\bibfield  {journal}
  {\bibinfo  {journal} {Phys. Rev. D}\ }\textbf {\bibinfo {volume} {100}},\
  \bibinfo {pages} {015043} (\bibinfo {year} {2019})},\ \Eprint
  {http://arxiv.org/abs/1812.03998} {arXiv:1812.03998 [hep-ph]} \BibitemShut
  {NoStop}%
\bibitem [{\citenamefont {Harnik}\ \emph {et~al.}(2019)\citenamefont {Harnik},
  \citenamefont {Liu},\ and\ \citenamefont {Palamara}}]{Harnik:2019zee}%
  \BibitemOpen
  \bibfield  {author} {\bibinfo {author} {\bibfnamefont {R.}~\bibnamefont
  {Harnik}}, \bibinfo {author} {\bibfnamefont {Z.}~\bibnamefont {Liu}}, \ and\
  \bibinfo {author} {\bibfnamefont {O.}~\bibnamefont {Palamara}},\ }\href
  {\doibase 10.1007/JHEP07(2019)170} {\bibfield  {journal} {\bibinfo  {journal}
  {JHEP}\ }\textbf {\bibinfo {volume} {07}},\ \bibinfo {pages} {170} (\bibinfo
  {year} {2019})},\ \Eprint {http://arxiv.org/abs/1902.03246} {arXiv:1902.03246
  [hep-ph]} \BibitemShut {NoStop}%
\bibitem [{\citenamefont {Davidson}\ \emph {et~al.}(2000)\citenamefont
  {Davidson}, \citenamefont {Hannestad},\ and\ \citenamefont
  {Raffelt}}]{Davidson:2000hf}%
  \BibitemOpen
  \bibfield  {author} {\bibinfo {author} {\bibfnamefont {S.}~\bibnamefont
  {Davidson}}, \bibinfo {author} {\bibfnamefont {S.}~\bibnamefont {Hannestad}},
  \ and\ \bibinfo {author} {\bibfnamefont {G.}~\bibnamefont {Raffelt}},\ }\href
  {\doibase 10.1088/1126-6708/2000/05/003} {\bibfield  {journal} {\bibinfo
  {journal} {JHEP}\ }\textbf {\bibinfo {volume} {05}},\ \bibinfo {pages} {003}
  (\bibinfo {year} {2000})},\ \Eprint {http://arxiv.org/abs/hep-ph/0001179}
  {arXiv:hep-ph/0001179} \BibitemShut {NoStop}%
\bibitem [{\citenamefont {Chang}\ \emph {et~al.}(2018)\citenamefont {Chang},
  \citenamefont {Essig},\ and\ \citenamefont {McDermott}}]{Chang:2018rso}%
  \BibitemOpen
  \bibfield  {author} {\bibinfo {author} {\bibfnamefont {J.~H.}\ \bibnamefont
  {Chang}}, \bibinfo {author} {\bibfnamefont {R.}~\bibnamefont {Essig}}, \ and\
  \bibinfo {author} {\bibfnamefont {S.~D.}\ \bibnamefont {McDermott}},\ }\href
  {\doibase 10.1007/JHEP09(2018)051} {\bibfield  {journal} {\bibinfo  {journal}
  {JHEP}\ }\textbf {\bibinfo {volume} {09}},\ \bibinfo {pages} {051} (\bibinfo
  {year} {2018})},\ \Eprint {http://arxiv.org/abs/1803.00993} {arXiv:1803.00993
  [hep-ph]} \BibitemShut {NoStop}%
\bibitem [{\citenamefont {Dvorkin}\ \emph {et~al.}(2019)\citenamefont
  {Dvorkin}, \citenamefont {Lin},\ and\ \citenamefont
  {Schutz}}]{Dvorkin:2019zdi}%
  \BibitemOpen
  \bibfield  {author} {\bibinfo {author} {\bibfnamefont {C.}~\bibnamefont
  {Dvorkin}}, \bibinfo {author} {\bibfnamefont {T.}~\bibnamefont {Lin}}, \ and\
  \bibinfo {author} {\bibfnamefont {K.}~\bibnamefont {Schutz}},\ }\href
  {\doibase 10.1103/PhysRevD.99.115009} {\bibfield  {journal} {\bibinfo
  {journal} {Phys. Rev. D}\ }\textbf {\bibinfo {volume} {99}},\ \bibinfo
  {pages} {115009} (\bibinfo {year} {2019})},\ \Eprint
  {http://arxiv.org/abs/1902.08623} {arXiv:1902.08623 [hep-ph]} \BibitemShut
  {NoStop}%
\bibitem [{\citenamefont {Creque-Sarbinowski}\ \emph
  {et~al.}(2019)\citenamefont {Creque-Sarbinowski}, \citenamefont {Ji},
  \citenamefont {Kovetz},\ and\ \citenamefont
  {Kamionkowski}}]{Creque-Sarbinowski:2019mcm}%
  \BibitemOpen
  \bibfield  {author} {\bibinfo {author} {\bibfnamefont {C.}~\bibnamefont
  {Creque-Sarbinowski}}, \bibinfo {author} {\bibfnamefont {L.}~\bibnamefont
  {Ji}}, \bibinfo {author} {\bibfnamefont {E.~D.}\ \bibnamefont {Kovetz}}, \
  and\ \bibinfo {author} {\bibfnamefont {M.}~\bibnamefont {Kamionkowski}},\
  }\href {\doibase 10.1103/PhysRevD.100.023528} {\bibfield  {journal} {\bibinfo
   {journal} {Phys. Rev. D}\ }\textbf {\bibinfo {volume} {100}},\ \bibinfo
  {pages} {023528} (\bibinfo {year} {2019})},\ \Eprint
  {http://arxiv.org/abs/1903.09154} {arXiv:1903.09154 [astro-ph.CO]}
  \BibitemShut {NoStop}%
\bibitem [{\citenamefont {Knapen}\ \emph {et~al.}(2018)\citenamefont {Knapen},
  \citenamefont {Lin}, \citenamefont {Pyle},\ and\ \citenamefont
  {Zurek}}]{Knapen:2017ekk}%
  \BibitemOpen
  \bibfield  {author} {\bibinfo {author} {\bibfnamefont {S.}~\bibnamefont
  {Knapen}}, \bibinfo {author} {\bibfnamefont {T.}~\bibnamefont {Lin}},
  \bibinfo {author} {\bibfnamefont {M.}~\bibnamefont {Pyle}}, \ and\ \bibinfo
  {author} {\bibfnamefont {K.~M.}\ \bibnamefont {Zurek}},\ }\href {\doibase
  10.1016/j.physletb.2018.08.064} {\bibfield  {journal} {\bibinfo  {journal}
  {Phys. Lett. B}\ }\textbf {\bibinfo {volume} {785}},\ \bibinfo {pages} {386}
  (\bibinfo {year} {2018})},\ \Eprint {http://arxiv.org/abs/1712.06598}
  {arXiv:1712.06598 [hep-ph]} \BibitemShut {NoStop}%
\bibitem [{\citenamefont {Blanco}\ \emph {et~al.}(2020)\citenamefont {Blanco},
  \citenamefont {Collar}, \citenamefont {Kahn},\ and\ \citenamefont
  {Lillard}}]{Blanco:2019lrf}%
  \BibitemOpen
  \bibfield  {author} {\bibinfo {author} {\bibfnamefont {C.}~\bibnamefont
  {Blanco}}, \bibinfo {author} {\bibfnamefont {J.~I.}\ \bibnamefont {Collar}},
  \bibinfo {author} {\bibfnamefont {Y.}~\bibnamefont {Kahn}}, \ and\ \bibinfo
  {author} {\bibfnamefont {B.}~\bibnamefont {Lillard}},\ }\href {\doibase
  10.1103/PhysRevD.101.056001} {\bibfield  {journal} {\bibinfo  {journal}
  {Phys. Rev. D}\ }\textbf {\bibinfo {volume} {101}},\ \bibinfo {pages}
  {056001} (\bibinfo {year} {2020})},\ \Eprint
  {http://arxiv.org/abs/1912.02822} {arXiv:1912.02822 [hep-ph]} \BibitemShut
  {NoStop}%
\bibitem [{\citenamefont {Essig}\ \emph {et~al.}(2019)\citenamefont {Essig},
  \citenamefont {P\'erez-R\'\i{}os}, \citenamefont {Ramani},\ and\
  \citenamefont {Slone}}]{Essig:2019kfe}%
  \BibitemOpen
  \bibfield  {author} {\bibinfo {author} {\bibfnamefont {R.}~\bibnamefont
  {Essig}}, \bibinfo {author} {\bibfnamefont {J.}~\bibnamefont
  {P\'erez-R\'\i{}os}}, \bibinfo {author} {\bibfnamefont {H.}~\bibnamefont
  {Ramani}}, \ and\ \bibinfo {author} {\bibfnamefont {O.}~\bibnamefont
  {Slone}},\ }\href {\doibase 10.1103/PhysRevResearch.1.033105} {\bibfield
  {journal} {\bibinfo  {journal} {Phys. Rev. Research.}\ }\textbf {\bibinfo
  {volume} {1}},\ \bibinfo {pages} {033105} (\bibinfo {year} {2019})},\ \Eprint
  {http://arxiv.org/abs/1907.07682} {arXiv:1907.07682 [hep-ph]} \BibitemShut
  {NoStop}%
\bibitem [{\citenamefont {Berlin}\ \emph {et~al.}(2020)\citenamefont {Berlin},
  \citenamefont {D'Agnolo}, \citenamefont {Ellis}, \citenamefont {Schuster},\
  and\ \citenamefont {Toro}}]{Berlin:2019uco}%
  \BibitemOpen
  \bibfield  {author} {\bibinfo {author} {\bibfnamefont {A.}~\bibnamefont
  {Berlin}}, \bibinfo {author} {\bibfnamefont {R.~T.}\ \bibnamefont
  {D'Agnolo}}, \bibinfo {author} {\bibfnamefont {S.~A.~R.}\ \bibnamefont
  {Ellis}}, \bibinfo {author} {\bibfnamefont {P.}~\bibnamefont {Schuster}}, \
  and\ \bibinfo {author} {\bibfnamefont {N.}~\bibnamefont {Toro}},\ }\href
  {\doibase 10.1103/PhysRevLett.124.011801} {\bibfield  {journal} {\bibinfo
  {journal} {Phys. Rev. Lett.}\ }\textbf {\bibinfo {volume} {124}},\ \bibinfo
  {pages} {011801} (\bibinfo {year} {2020})},\ \Eprint
  {http://arxiv.org/abs/1908.06982} {arXiv:1908.06982 [hep-ph]} \BibitemShut
  {NoStop}%
\bibitem [{\citenamefont {Kurinsky}\ \emph {et~al.}(2019)\citenamefont
  {Kurinsky}, \citenamefont {Yu}, \citenamefont {Hochberg},\ and\ \citenamefont
  {Cabrera}}]{Kurinsky:2019pgb}%
  \BibitemOpen
  \bibfield  {author} {\bibinfo {author} {\bibfnamefont {N.~A.}\ \bibnamefont
  {Kurinsky}}, \bibinfo {author} {\bibfnamefont {T.~C.}\ \bibnamefont {Yu}},
  \bibinfo {author} {\bibfnamefont {Y.}~\bibnamefont {Hochberg}}, \ and\
  \bibinfo {author} {\bibfnamefont {B.}~\bibnamefont {Cabrera}},\ }\href
  {\doibase 10.1103/PhysRevD.99.123005} {\bibfield  {journal} {\bibinfo
  {journal} {Phys. Rev. D}\ }\textbf {\bibinfo {volume} {99}},\ \bibinfo
  {pages} {123005} (\bibinfo {year} {2019})},\ \Eprint
  {http://arxiv.org/abs/1901.07569} {arXiv:1901.07569 [hep-ex]} \BibitemShut
  {NoStop}%
\bibitem [{\citenamefont {Barak}\ \emph {et~al.}(2020)\citenamefont {Barak}
  \emph {et~al.}}]{Barak:2020fql}%
  \BibitemOpen
  \bibfield  {author} {\bibinfo {author} {\bibfnamefont {L.}~\bibnamefont
  {Barak}} \emph {et~al.} (\bibinfo {collaboration} {SENSEI}),\ }\href
  {\doibase 10.1103/PhysRevLett.125.171802} {\bibfield  {journal} {\bibinfo
  {journal} {Phys. Rev. Lett.}\ }\textbf {\bibinfo {volume} {125}},\ \bibinfo
  {pages} {171802} (\bibinfo {year} {2020})},\ \Eprint
  {http://arxiv.org/abs/2004.11378} {arXiv:2004.11378 [astro-ph.CO]}
  \BibitemShut {NoStop}%
\bibitem [{\citenamefont {Griffin}\ \emph {et~al.}(2021)\citenamefont
  {Griffin}, \citenamefont {Hochberg}, \citenamefont {Inzani}, \citenamefont
  {Kurinsky}, \citenamefont {Lin},\ and\ \citenamefont
  {Chin}}]{Griffin:2020lgd}%
  \BibitemOpen
  \bibfield  {author} {\bibinfo {author} {\bibfnamefont {S.~M.}\ \bibnamefont
  {Griffin}}, \bibinfo {author} {\bibfnamefont {Y.}~\bibnamefont {Hochberg}},
  \bibinfo {author} {\bibfnamefont {K.}~\bibnamefont {Inzani}}, \bibinfo
  {author} {\bibfnamefont {N.}~\bibnamefont {Kurinsky}}, \bibinfo {author}
  {\bibfnamefont {T.}~\bibnamefont {Lin}}, \ and\ \bibinfo {author}
  {\bibfnamefont {T.}~\bibnamefont {Chin}},\ }\href {\doibase
  10.1103/PhysRevD.103.075002} {\bibfield  {journal} {\bibinfo  {journal}
  {Phys. Rev. D}\ }\textbf {\bibinfo {volume} {103}},\ \bibinfo {pages}
  {075002} (\bibinfo {year} {2021})},\ \Eprint
  {http://arxiv.org/abs/2008.08560} {arXiv:2008.08560 [hep-ph]} \BibitemShut
  {NoStop}%
\bibitem [{\citenamefont {Barkana}\ \emph {et~al.}(2018)\citenamefont
  {Barkana}, \citenamefont {Outmezguine}, \citenamefont {Redigolo},\ and\
  \citenamefont {Volansky}}]{Barkana:2018qrx}%
  \BibitemOpen
  \bibfield  {author} {\bibinfo {author} {\bibfnamefont {R.}~\bibnamefont
  {Barkana}}, \bibinfo {author} {\bibfnamefont {N.~J.}\ \bibnamefont
  {Outmezguine}}, \bibinfo {author} {\bibfnamefont {D.}~\bibnamefont
  {Redigolo}}, \ and\ \bibinfo {author} {\bibfnamefont {T.}~\bibnamefont
  {Volansky}},\ }\href {\doibase 10.1103/PhysRevD.98.103005} {\bibfield
  {journal} {\bibinfo  {journal} {Phys. Rev. D}\ }\textbf {\bibinfo {volume}
  {98}},\ \bibinfo {pages} {103005} (\bibinfo {year} {2018})},\ \Eprint
  {http://arxiv.org/abs/1803.03091} {arXiv:1803.03091 [hep-ph]} \BibitemShut
  {NoStop}%
\bibitem [{\citenamefont {Mu\~noz}\ and\ \citenamefont
  {Loeb}(2018)}]{Munoz:2018pzp}%
  \BibitemOpen
  \bibfield  {author} {\bibinfo {author} {\bibfnamefont {J.~B.}\ \bibnamefont
  {Mu\~noz}}\ and\ \bibinfo {author} {\bibfnamefont {A.}~\bibnamefont {Loeb}},\
  }\href {\doibase 10.1038/s41586-018-0151-x} {\bibfield  {journal} {\bibinfo
  {journal} {Nature}\ }\textbf {\bibinfo {volume} {557}},\ \bibinfo {pages}
  {684} (\bibinfo {year} {2018})},\ \Eprint {http://arxiv.org/abs/1802.10094}
  {arXiv:1802.10094 [astro-ph.CO]} \BibitemShut {NoStop}%
\bibitem [{\citenamefont {Liu}\ \emph {et~al.}(2019)\citenamefont {Liu},
  \citenamefont {Outmezguine}, \citenamefont {Redigolo},\ and\ \citenamefont
  {Volansky}}]{Liu:2019knx}%
  \BibitemOpen
  \bibfield  {author} {\bibinfo {author} {\bibfnamefont {H.}~\bibnamefont
  {Liu}}, \bibinfo {author} {\bibfnamefont {N.~J.}\ \bibnamefont
  {Outmezguine}}, \bibinfo {author} {\bibfnamefont {D.}~\bibnamefont
  {Redigolo}}, \ and\ \bibinfo {author} {\bibfnamefont {T.}~\bibnamefont
  {Volansky}},\ }\href {\doibase 10.1103/PhysRevD.100.123011} {\bibfield
  {journal} {\bibinfo  {journal} {Phys. Rev. D}\ }\textbf {\bibinfo {volume}
  {100}},\ \bibinfo {pages} {123011} (\bibinfo {year} {2019})},\ \Eprint
  {http://arxiv.org/abs/1908.06986} {arXiv:1908.06986 [hep-ph]} \BibitemShut
  {NoStop}%
\bibitem [{\citenamefont {Kurinsky}\ \emph {et~al.}(2020)\citenamefont
  {Kurinsky}, \citenamefont {Baxter}, \citenamefont {Kahn},\ and\ \citenamefont
  {Krnjaic}}]{Kurinsky:2020dpb}%
  \BibitemOpen
  \bibfield  {author} {\bibinfo {author} {\bibfnamefont {N.}~\bibnamefont
  {Kurinsky}}, \bibinfo {author} {\bibfnamefont {D.}~\bibnamefont {Baxter}},
  \bibinfo {author} {\bibfnamefont {Y.}~\bibnamefont {Kahn}}, \ and\ \bibinfo
  {author} {\bibfnamefont {G.}~\bibnamefont {Krnjaic}},\ }\href {\doibase
  10.1103/PhysRevD.102.015017} {\bibfield  {journal} {\bibinfo  {journal}
  {Phys. Rev. D}\ }\textbf {\bibinfo {volume} {102}},\ \bibinfo {pages}
  {015017} (\bibinfo {year} {2020})},\ \Eprint
  {http://arxiv.org/abs/2002.06937} {arXiv:2002.06937 [hep-ph]} \BibitemShut
  {NoStop}%
\bibitem [{\citenamefont {Bloch}\ \emph {et~al.}(2021)\citenamefont {Bloch},
  \citenamefont {Caputo}, \citenamefont {Essig}, \citenamefont {Redigolo},
  \citenamefont {Sholapurkar},\ and\ \citenamefont {Volansky}}]{Bloch:2020uzh}%
  \BibitemOpen
  \bibfield  {author} {\bibinfo {author} {\bibfnamefont {I.~M.}\ \bibnamefont
  {Bloch}}, \bibinfo {author} {\bibfnamefont {A.}~\bibnamefont {Caputo}},
  \bibinfo {author} {\bibfnamefont {R.}~\bibnamefont {Essig}}, \bibinfo
  {author} {\bibfnamefont {D.}~\bibnamefont {Redigolo}}, \bibinfo {author}
  {\bibfnamefont {M.}~\bibnamefont {Sholapurkar}}, \ and\ \bibinfo {author}
  {\bibfnamefont {T.}~\bibnamefont {Volansky}},\ }\href {\doibase
  10.1007/JHEP01(2021)178} {\bibfield  {journal} {\bibinfo  {journal} {JHEP}\
  }\textbf {\bibinfo {volume} {01}},\ \bibinfo {pages} {178} (\bibinfo {year}
  {2021})},\ \Eprint {http://arxiv.org/abs/2006.14521} {arXiv:2006.14521
  [hep-ph]} \BibitemShut {NoStop}%
\bibitem [{\citenamefont {Harnik}\ \emph {et~al.}(2021)\citenamefont {Harnik},
  \citenamefont {Plestid}, \citenamefont {Pospelov},\ and\ \citenamefont
  {Ramani}}]{Harnik:2020ugb}%
  \BibitemOpen
  \bibfield  {author} {\bibinfo {author} {\bibfnamefont {R.}~\bibnamefont
  {Harnik}}, \bibinfo {author} {\bibfnamefont {R.}~\bibnamefont {Plestid}},
  \bibinfo {author} {\bibfnamefont {M.}~\bibnamefont {Pospelov}}, \ and\
  \bibinfo {author} {\bibfnamefont {H.}~\bibnamefont {Ramani}},\ }\href
  {\doibase 10.1103/PhysRevD.103.075029} {\bibfield  {journal} {\bibinfo
  {journal} {Phys. Rev. D}\ }\textbf {\bibinfo {volume} {103}},\ \bibinfo
  {pages} {075029} (\bibinfo {year} {2021})},\ \Eprint
  {http://arxiv.org/abs/2010.11190} {arXiv:2010.11190 [hep-ph]} \BibitemShut
  {NoStop}%
\bibitem [{\citenamefont {Emken}\ \emph {et~al.}(2019)\citenamefont {Emken},
  \citenamefont {Essig}, \citenamefont {Kouvaris},\ and\ \citenamefont
  {Sholapurkar}}]{Emken:2019tni}%
  \BibitemOpen
  \bibfield  {author} {\bibinfo {author} {\bibfnamefont {T.}~\bibnamefont
  {Emken}}, \bibinfo {author} {\bibfnamefont {R.}~\bibnamefont {Essig}},
  \bibinfo {author} {\bibfnamefont {C.}~\bibnamefont {Kouvaris}}, \ and\
  \bibinfo {author} {\bibfnamefont {M.}~\bibnamefont {Sholapurkar}},\ }\href
  {\doibase 10.1088/1475-7516/2019/09/070} {\bibfield  {journal} {\bibinfo
  {journal} {JCAP}\ }\textbf {\bibinfo {volume} {09}},\ \bibinfo {pages} {070}
  (\bibinfo {year} {2019})},\ \Eprint {http://arxiv.org/abs/1905.06348}
  {arXiv:1905.06348 [hep-ph]} \BibitemShut {NoStop}%
\bibitem [{\citenamefont {Pospelov}\ and\ \citenamefont
  {Ramani}(2021)}]{Pospelov:2020ktu}%
  \BibitemOpen
  \bibfield  {author} {\bibinfo {author} {\bibfnamefont {M.}~\bibnamefont
  {Pospelov}}\ and\ \bibinfo {author} {\bibfnamefont {H.}~\bibnamefont
  {Ramani}},\ }\href {\doibase 10.1103/PhysRevD.103.115031} {\bibfield
  {journal} {\bibinfo  {journal} {Phys. Rev. D}\ }\textbf {\bibinfo {volume}
  {103}},\ \bibinfo {pages} {115031} (\bibinfo {year} {2021})},\ \Eprint
  {http://arxiv.org/abs/2012.03957} {arXiv:2012.03957 [hep-ph]} \BibitemShut
  {NoStop}%
\bibitem [{\citenamefont {Kim}\ \emph {et~al.}(2007)\citenamefont {Kim},
  \citenamefont {Lee}, \citenamefont {Lee}, \citenamefont {Perl}, \citenamefont
  {Halyo},\ and\ \citenamefont {Loomba}}]{kim2007search}%
  \BibitemOpen
  \bibfield  {author} {\bibinfo {author} {\bibfnamefont {P.~C.}\ \bibnamefont
  {Kim}}, \bibinfo {author} {\bibfnamefont {E.~R.}\ \bibnamefont {Lee}},
  \bibinfo {author} {\bibfnamefont {I.~T.}\ \bibnamefont {Lee}}, \bibinfo
  {author} {\bibfnamefont {M.~L.}\ \bibnamefont {Perl}}, \bibinfo {author}
  {\bibfnamefont {V.}~\bibnamefont {Halyo}}, \ and\ \bibinfo {author}
  {\bibfnamefont {D.}~\bibnamefont {Loomba}},\ }\href@noop {} {\bibfield
  {journal} {\bibinfo  {journal} {Physical review letters}\ }\textbf {\bibinfo
  {volume} {99}},\ \bibinfo {pages} {161804} (\bibinfo {year}
  {2007})}\BibitemShut {NoStop}%
\bibitem [{\citenamefont {Moore}\ \emph {et~al.}(2014)\citenamefont {Moore},
  \citenamefont {Rider},\ and\ \citenamefont {Gratta}}]{moore2014search}%
  \BibitemOpen
  \bibfield  {author} {\bibinfo {author} {\bibfnamefont {D.~C.}\ \bibnamefont
  {Moore}}, \bibinfo {author} {\bibfnamefont {A.~D.}\ \bibnamefont {Rider}}, \
  and\ \bibinfo {author} {\bibfnamefont {G.}~\bibnamefont {Gratta}},\
  }\href@noop {} {\bibfield  {journal} {\bibinfo  {journal} {Physical review
  letters}\ }\textbf {\bibinfo {volume} {113}},\ \bibinfo {pages} {251801}
  (\bibinfo {year} {2014})}\BibitemShut {NoStop}%
\bibitem [{\citenamefont {Afek}\ \emph {et~al.}(2020)\citenamefont {Afek},
  \citenamefont {Monteiro}, \citenamefont {Wang}, \citenamefont {Siegel},
  \citenamefont {Ghosh},\ and\ \citenamefont {Moore}}]{afek2020limits}%
  \BibitemOpen
  \bibfield  {author} {\bibinfo {author} {\bibfnamefont {G.}~\bibnamefont
  {Afek}}, \bibinfo {author} {\bibfnamefont {F.}~\bibnamefont {Monteiro}},
  \bibinfo {author} {\bibfnamefont {J.}~\bibnamefont {Wang}}, \bibinfo {author}
  {\bibfnamefont {B.}~\bibnamefont {Siegel}}, \bibinfo {author} {\bibfnamefont
  {S.}~\bibnamefont {Ghosh}}, \ and\ \bibinfo {author} {\bibfnamefont {D.~C.}\
  \bibnamefont {Moore}},\ }\href@noop {} {\bibfield  {journal} {\bibinfo
  {journal} {arXiv preprint arXiv:2012.08169}\ } (\bibinfo {year}
  {2020})}\BibitemShut {NoStop}%
\bibitem [{\citenamefont {Hanneke}\ \emph {et~al.}(2011)\citenamefont
  {Hanneke}, \citenamefont {Hoogerheide},\ and\ \citenamefont
  {Gabrielse}}]{hanneke2011cavity}%
  \BibitemOpen
  \bibfield  {author} {\bibinfo {author} {\bibfnamefont {D.}~\bibnamefont
  {Hanneke}}, \bibinfo {author} {\bibfnamefont {S.~F.}\ \bibnamefont
  {Hoogerheide}}, \ and\ \bibinfo {author} {\bibfnamefont {G.}~\bibnamefont
  {Gabrielse}},\ }\href@noop {} {\bibfield  {journal} {\bibinfo  {journal}
  {Physical Review A}\ }\textbf {\bibinfo {volume} {83}},\ \bibinfo {pages}
  {052122} (\bibinfo {year} {2011})}\BibitemShut {NoStop}%
\bibitem [{\citenamefont {Cairncross}\ \emph {et~al.}(2017)\citenamefont
  {Cairncross}, \citenamefont {Gresh}, \citenamefont {Grau}, \citenamefont
  {Cossel}, \citenamefont {Roussy}, \citenamefont {Ni}, \citenamefont {Zhou},
  \citenamefont {Ye},\ and\ \citenamefont {Cornell}}]{cairncross2017precision}%
  \BibitemOpen
  \bibfield  {author} {\bibinfo {author} {\bibfnamefont {W.~B.}\ \bibnamefont
  {Cairncross}}, \bibinfo {author} {\bibfnamefont {D.~N.}\ \bibnamefont
  {Gresh}}, \bibinfo {author} {\bibfnamefont {M.}~\bibnamefont {Grau}},
  \bibinfo {author} {\bibfnamefont {K.~C.}\ \bibnamefont {Cossel}}, \bibinfo
  {author} {\bibfnamefont {T.~S.}\ \bibnamefont {Roussy}}, \bibinfo {author}
  {\bibfnamefont {Y.}~\bibnamefont {Ni}}, \bibinfo {author} {\bibfnamefont
  {Y.}~\bibnamefont {Zhou}}, \bibinfo {author} {\bibfnamefont {J.}~\bibnamefont
  {Ye}}, \ and\ \bibinfo {author} {\bibfnamefont {E.~A.}\ \bibnamefont
  {Cornell}},\ }\href@noop {} {\bibfield  {journal} {\bibinfo  {journal}
  {Physical review letters}\ }\textbf {\bibinfo {volume} {119}},\ \bibinfo
  {pages} {153001} (\bibinfo {year} {2017})}\BibitemShut {NoStop}%
\bibitem [{\citenamefont {Schneider}\ \emph {et~al.}(2017)\citenamefont
  {Schneider}, \citenamefont {Mooser}, \citenamefont {Bohman}, \citenamefont
  {Sch{\"o}n}, \citenamefont {Harrington}, \citenamefont {Higuchi},
  \citenamefont {Nagahama}, \citenamefont {Sellner}, \citenamefont {Smorra},
  \citenamefont {Blaum} \emph {et~al.}}]{schneider2017double}%
  \BibitemOpen
  \bibfield  {author} {\bibinfo {author} {\bibfnamefont {G.}~\bibnamefont
  {Schneider}}, \bibinfo {author} {\bibfnamefont {A.}~\bibnamefont {Mooser}},
  \bibinfo {author} {\bibfnamefont {M.}~\bibnamefont {Bohman}}, \bibinfo
  {author} {\bibfnamefont {N.}~\bibnamefont {Sch{\"o}n}}, \bibinfo {author}
  {\bibfnamefont {J.}~\bibnamefont {Harrington}}, \bibinfo {author}
  {\bibfnamefont {T.}~\bibnamefont {Higuchi}}, \bibinfo {author} {\bibfnamefont
  {H.}~\bibnamefont {Nagahama}}, \bibinfo {author} {\bibfnamefont
  {S.}~\bibnamefont {Sellner}}, \bibinfo {author} {\bibfnamefont
  {C.}~\bibnamefont {Smorra}}, \bibinfo {author} {\bibfnamefont
  {K.}~\bibnamefont {Blaum}},  \emph {et~al.},\ }\href@noop {} {\bibfield
  {journal} {\bibinfo  {journal} {Science}\ }\textbf {\bibinfo {volume}
  {358}},\ \bibinfo {pages} {1081} (\bibinfo {year} {2017})}\BibitemShut
  {NoStop}%
\bibitem [{\citenamefont {Smorra}\ \emph {et~al.}(2017)\citenamefont {Smorra},
  \citenamefont {Sellner}, \citenamefont {Borchert}, \citenamefont
  {Harrington}, \citenamefont {Higuchi}, \citenamefont {Nagahama},
  \citenamefont {Tanaka}, \citenamefont {Mooser}, \citenamefont {Schneider},
  \citenamefont {Bohman} \emph {et~al.}}]{smorra2017parts}%
  \BibitemOpen
  \bibfield  {author} {\bibinfo {author} {\bibfnamefont {C.}~\bibnamefont
  {Smorra}}, \bibinfo {author} {\bibfnamefont {S.}~\bibnamefont {Sellner}},
  \bibinfo {author} {\bibfnamefont {M.}~\bibnamefont {Borchert}}, \bibinfo
  {author} {\bibfnamefont {J.}~\bibnamefont {Harrington}}, \bibinfo {author}
  {\bibfnamefont {T.}~\bibnamefont {Higuchi}}, \bibinfo {author} {\bibfnamefont
  {H.}~\bibnamefont {Nagahama}}, \bibinfo {author} {\bibfnamefont
  {T.}~\bibnamefont {Tanaka}}, \bibinfo {author} {\bibfnamefont
  {A.}~\bibnamefont {Mooser}}, \bibinfo {author} {\bibfnamefont
  {G.}~\bibnamefont {Schneider}}, \bibinfo {author} {\bibfnamefont
  {M.}~\bibnamefont {Bohman}},  \emph {et~al.},\ }\href@noop {} {\bibfield
  {journal} {\bibinfo  {journal} {Nature}\ }\textbf {\bibinfo {volume} {550}},\
  \bibinfo {pages} {371} (\bibinfo {year} {2017})}\BibitemShut {NoStop}%
\bibitem [{\citenamefont {Roberts}\ \emph {et~al.}(2020)\citenamefont
  {Roberts}, \citenamefont {Delva}, \citenamefont {Al-Masoudi}, \citenamefont
  {Amy-Klein}, \citenamefont {B{\ae}rentsen}, \citenamefont {Baynham},
  \citenamefont {Benkler}, \citenamefont {Bilicki}, \citenamefont {Bize},
  \citenamefont {Bowden}, \citenamefont {Calvert}, \citenamefont {Cambier},
  \citenamefont {Cantin}, \citenamefont {Curtis}, \citenamefont {D{\"o}rscher},
  \citenamefont {Favier}, \citenamefont {Frank}, \citenamefont {Gill},
  \citenamefont {Godun}, \citenamefont {Grosche}, \citenamefont {Guo},
  \citenamefont {Hees}, \citenamefont {Hill}, \citenamefont {Hobson},
  \citenamefont {Huntemann}, \citenamefont {Kronj{\"a}ger}, \citenamefont
  {Koke}, \citenamefont {Kuhl}, \citenamefont {Lange}, \citenamefont {Legero},
  \citenamefont {Lipphardt}, \citenamefont {Lisdat}, \citenamefont {Lodewyck},
  \citenamefont {Lopez}, \citenamefont {Margolis}, \citenamefont
  {{\'{A}}lvarez-Mart{\'{\i}}nez}, \citenamefont {Meynadier}, \citenamefont
  {Ozimek}, \citenamefont {Peik}, \citenamefont {Pottie}, \citenamefont
  {Quintin}, \citenamefont {Sanner}, \citenamefont {Sarlo}, \citenamefont
  {Schioppo}, \citenamefont {Schwarz}, \citenamefont {Silva}, \citenamefont
  {Sterr}, \citenamefont {Tamm}, \citenamefont {Targat}, \citenamefont
  {Tuckey}, \citenamefont {Vallet}, \citenamefont {Waterholter}, \citenamefont
  {Xu},\ and\ \citenamefont {Wolf}}]{Roberts_2020}%
  \BibitemOpen
  \bibfield  {author} {\bibinfo {author} {\bibfnamefont {B.~M.}\ \bibnamefont
  {Roberts}}, \bibinfo {author} {\bibfnamefont {P.}~\bibnamefont {Delva}},
  \bibinfo {author} {\bibfnamefont {A.}~\bibnamefont {Al-Masoudi}}, \bibinfo
  {author} {\bibfnamefont {A.}~\bibnamefont {Amy-Klein}}, \bibinfo {author}
  {\bibfnamefont {C.}~\bibnamefont {B{\ae}rentsen}}, \bibinfo {author}
  {\bibfnamefont {C.~F.~A.}\ \bibnamefont {Baynham}}, \bibinfo {author}
  {\bibfnamefont {E.}~\bibnamefont {Benkler}}, \bibinfo {author} {\bibfnamefont
  {S.}~\bibnamefont {Bilicki}}, \bibinfo {author} {\bibfnamefont
  {S.}~\bibnamefont {Bize}}, \bibinfo {author} {\bibfnamefont {W.}~\bibnamefont
  {Bowden}}, \bibinfo {author} {\bibfnamefont {J.}~\bibnamefont {Calvert}},
  \bibinfo {author} {\bibfnamefont {V.}~\bibnamefont {Cambier}}, \bibinfo
  {author} {\bibfnamefont {E.}~\bibnamefont {Cantin}}, \bibinfo {author}
  {\bibfnamefont {E.~A.}\ \bibnamefont {Curtis}}, \bibinfo {author}
  {\bibfnamefont {S.}~\bibnamefont {D{\"o}rscher}}, \bibinfo {author}
  {\bibfnamefont {M.}~\bibnamefont {Favier}}, \bibinfo {author} {\bibfnamefont
  {F.}~\bibnamefont {Frank}}, \bibinfo {author} {\bibfnamefont
  {P.}~\bibnamefont {Gill}}, \bibinfo {author} {\bibfnamefont {R.~M.}\
  \bibnamefont {Godun}}, \bibinfo {author} {\bibfnamefont {G.}~\bibnamefont
  {Grosche}}, \bibinfo {author} {\bibfnamefont {C.}~\bibnamefont {Guo}},
  \bibinfo {author} {\bibfnamefont {A.}~\bibnamefont {Hees}}, \bibinfo {author}
  {\bibfnamefont {I.~R.}\ \bibnamefont {Hill}}, \bibinfo {author}
  {\bibfnamefont {R.}~\bibnamefont {Hobson}}, \bibinfo {author} {\bibfnamefont
  {N.}~\bibnamefont {Huntemann}}, \bibinfo {author} {\bibfnamefont
  {J.}~\bibnamefont {Kronj{\"a}ger}}, \bibinfo {author} {\bibfnamefont
  {S.}~\bibnamefont {Koke}}, \bibinfo {author} {\bibfnamefont {A.}~\bibnamefont
  {Kuhl}}, \bibinfo {author} {\bibfnamefont {R.}~\bibnamefont {Lange}},
  \bibinfo {author} {\bibfnamefont {T.}~\bibnamefont {Legero}}, \bibinfo
  {author} {\bibfnamefont {B.}~\bibnamefont {Lipphardt}}, \bibinfo {author}
  {\bibfnamefont {C.}~\bibnamefont {Lisdat}}, \bibinfo {author} {\bibfnamefont
  {J.}~\bibnamefont {Lodewyck}}, \bibinfo {author} {\bibfnamefont
  {O.}~\bibnamefont {Lopez}}, \bibinfo {author} {\bibfnamefont {H.~S.}\
  \bibnamefont {Margolis}}, \bibinfo {author} {\bibfnamefont {H.}~\bibnamefont
  {{\'{A}}lvarez-Mart{\'{\i}}nez}}, \bibinfo {author} {\bibfnamefont
  {F.}~\bibnamefont {Meynadier}}, \bibinfo {author} {\bibfnamefont
  {F.}~\bibnamefont {Ozimek}}, \bibinfo {author} {\bibfnamefont
  {E.}~\bibnamefont {Peik}}, \bibinfo {author} {\bibfnamefont {P.-E.}\
  \bibnamefont {Pottie}}, \bibinfo {author} {\bibfnamefont {N.}~\bibnamefont
  {Quintin}}, \bibinfo {author} {\bibfnamefont {C.}~\bibnamefont {Sanner}},
  \bibinfo {author} {\bibfnamefont {L.~D.}\ \bibnamefont {Sarlo}}, \bibinfo
  {author} {\bibfnamefont {M.}~\bibnamefont {Schioppo}}, \bibinfo {author}
  {\bibfnamefont {R.}~\bibnamefont {Schwarz}}, \bibinfo {author} {\bibfnamefont
  {A.}~\bibnamefont {Silva}}, \bibinfo {author} {\bibfnamefont
  {U.}~\bibnamefont {Sterr}}, \bibinfo {author} {\bibfnamefont
  {C.}~\bibnamefont {Tamm}}, \bibinfo {author} {\bibfnamefont {R.~L.}\
  \bibnamefont {Targat}}, \bibinfo {author} {\bibfnamefont {P.}~\bibnamefont
  {Tuckey}}, \bibinfo {author} {\bibfnamefont {G.}~\bibnamefont {Vallet}},
  \bibinfo {author} {\bibfnamefont {T.}~\bibnamefont {Waterholter}}, \bibinfo
  {author} {\bibfnamefont {D.}~\bibnamefont {Xu}}, \ and\ \bibinfo {author}
  {\bibfnamefont {P.}~\bibnamefont {Wolf}},\ }\href {\doibase
  10.1088/1367-2630/abaace} {\bibfield  {journal} {\bibinfo  {journal} {New
  Journal of Physics}\ }\textbf {\bibinfo {volume} {22}},\ \bibinfo {pages}
  {093010} (\bibinfo {year} {2020})}\BibitemShut {NoStop}%
\bibitem [{\citenamefont {H{\"a}ffner}\ \emph {et~al.}(2008)\citenamefont
  {H{\"a}ffner}, \citenamefont {Roos},\ and\ \citenamefont
  {Blatt}}]{haffner2008quantum}%
  \BibitemOpen
  \bibfield  {author} {\bibinfo {author} {\bibfnamefont {H.}~\bibnamefont
  {H{\"a}ffner}}, \bibinfo {author} {\bibfnamefont {C.~F.}\ \bibnamefont
  {Roos}}, \ and\ \bibinfo {author} {\bibfnamefont {R.}~\bibnamefont {Blatt}},\
  }\href@noop {} {\bibfield  {journal} {\bibinfo  {journal} {Physics reports}\
  }\textbf {\bibinfo {volume} {469}},\ \bibinfo {pages} {155} (\bibinfo {year}
  {2008})}\BibitemShut {NoStop}%
\bibitem [{\citenamefont {Hite}\ \emph {et~al.}(2012)\citenamefont {Hite},
  \citenamefont {Colombe}, \citenamefont {Wilson}, \citenamefont {Brown},
  \citenamefont {Warring}, \citenamefont {J{\"o}rdens}, \citenamefont {Jost},
  \citenamefont {McKay}, \citenamefont {Pappas}, \citenamefont {Leibfried}
  \emph {et~al.}}]{hite2012100}%
  \BibitemOpen
  \bibfield  {author} {\bibinfo {author} {\bibfnamefont {D.~A.}\ \bibnamefont
  {Hite}}, \bibinfo {author} {\bibfnamefont {Y.}~\bibnamefont {Colombe}},
  \bibinfo {author} {\bibfnamefont {A.~C.}\ \bibnamefont {Wilson}}, \bibinfo
  {author} {\bibfnamefont {K.~R.}\ \bibnamefont {Brown}}, \bibinfo {author}
  {\bibfnamefont {U.}~\bibnamefont {Warring}}, \bibinfo {author} {\bibfnamefont
  {R.}~\bibnamefont {J{\"o}rdens}}, \bibinfo {author} {\bibfnamefont {J.~D.}\
  \bibnamefont {Jost}}, \bibinfo {author} {\bibfnamefont {K.}~\bibnamefont
  {McKay}}, \bibinfo {author} {\bibfnamefont {D.}~\bibnamefont {Pappas}},
  \bibinfo {author} {\bibfnamefont {D.}~\bibnamefont {Leibfried}},  \emph
  {et~al.},\ }\href@noop {} {\bibfield  {journal} {\bibinfo  {journal}
  {Physical review letters}\ }\textbf {\bibinfo {volume} {109}},\ \bibinfo
  {pages} {103001} (\bibinfo {year} {2012})}\BibitemShut {NoStop}%
\bibitem [{\citenamefont {Safronova}\ \emph {et~al.}(2018)\citenamefont
  {Safronova}, \citenamefont {Budker}, \citenamefont {DeMille}, \citenamefont
  {Kimball}, \citenamefont {Derevianko},\ and\ \citenamefont
  {Clark}}]{safronova2018search}%
  \BibitemOpen
  \bibfield  {author} {\bibinfo {author} {\bibfnamefont {M.}~\bibnamefont
  {Safronova}}, \bibinfo {author} {\bibfnamefont {D.}~\bibnamefont {Budker}},
  \bibinfo {author} {\bibfnamefont {D.}~\bibnamefont {DeMille}}, \bibinfo
  {author} {\bibfnamefont {D.~F.~J.}\ \bibnamefont {Kimball}}, \bibinfo
  {author} {\bibfnamefont {A.}~\bibnamefont {Derevianko}}, \ and\ \bibinfo
  {author} {\bibfnamefont {C.~W.}\ \bibnamefont {Clark}},\ }\href@noop {}
  {\bibfield  {journal} {\bibinfo  {journal} {Reviews of Modern Physics}\
  }\textbf {\bibinfo {volume} {90}},\ \bibinfo {pages} {025008} (\bibinfo
  {year} {2018})}\BibitemShut {NoStop}%
\bibitem [{\citenamefont {Sellner}\ \emph {et~al.}(2017)\citenamefont
  {Sellner}, \citenamefont {Besirli}, \citenamefont {Bohman}, \citenamefont
  {Borchert}, \citenamefont {Harrington}, \citenamefont {Higuchi},
  \citenamefont {Mooser}, \citenamefont {Nagahama}, \citenamefont {Schneider},
  \citenamefont {Smorra} \emph {et~al.}}]{sellner2017improved}%
  \BibitemOpen
  \bibfield  {author} {\bibinfo {author} {\bibfnamefont {S.}~\bibnamefont
  {Sellner}}, \bibinfo {author} {\bibfnamefont {M.}~\bibnamefont {Besirli}},
  \bibinfo {author} {\bibfnamefont {M.}~\bibnamefont {Bohman}}, \bibinfo
  {author} {\bibfnamefont {M.}~\bibnamefont {Borchert}}, \bibinfo {author}
  {\bibfnamefont {J.}~\bibnamefont {Harrington}}, \bibinfo {author}
  {\bibfnamefont {T.}~\bibnamefont {Higuchi}}, \bibinfo {author} {\bibfnamefont
  {A.}~\bibnamefont {Mooser}}, \bibinfo {author} {\bibfnamefont
  {H.}~\bibnamefont {Nagahama}}, \bibinfo {author} {\bibfnamefont
  {G.}~\bibnamefont {Schneider}}, \bibinfo {author} {\bibfnamefont
  {C.}~\bibnamefont {Smorra}},  \emph {et~al.},\ }\href@noop {} {\bibfield
  {journal} {\bibinfo  {journal} {New Journal of Physics}\ }\textbf {\bibinfo
  {volume} {19}},\ \bibinfo {pages} {083023} (\bibinfo {year}
  {2017})}\BibitemShut {NoStop}%
\bibitem [{\citenamefont {Goodwin}\ \emph {et~al.}(2016)\citenamefont
  {Goodwin}, \citenamefont {Stutter}, \citenamefont {Thompson},\ and\
  \citenamefont {Segal}}]{goodwin2016resolved}%
  \BibitemOpen
  \bibfield  {author} {\bibinfo {author} {\bibfnamefont {J.~F.}\ \bibnamefont
  {Goodwin}}, \bibinfo {author} {\bibfnamefont {G.}~\bibnamefont {Stutter}},
  \bibinfo {author} {\bibfnamefont {R.~C.}\ \bibnamefont {Thompson}}, \ and\
  \bibinfo {author} {\bibfnamefont {D.~M.}\ \bibnamefont {Segal}},\ }\href@noop
  {} {\bibfield  {journal} {\bibinfo  {journal} {Physical review letters}\
  }\textbf {\bibinfo {volume} {116}},\ \bibinfo {pages} {143002} (\bibinfo
  {year} {2016})}\BibitemShut {NoStop}%
\bibitem [{\citenamefont {Borchert}\ \emph {et~al.}(2019)\citenamefont
  {Borchert}, \citenamefont {Blessing}, \citenamefont {Devlin}, \citenamefont
  {Harrington}, \citenamefont {Higuchi}, \citenamefont {Morgner}, \citenamefont
  {Smorra}, \citenamefont {Wursten}, \citenamefont {Bohman}, \citenamefont
  {Wiesinger} \emph {et~al.}}]{borchert2019measurement}%
  \BibitemOpen
  \bibfield  {author} {\bibinfo {author} {\bibfnamefont {M.}~\bibnamefont
  {Borchert}}, \bibinfo {author} {\bibfnamefont {P.}~\bibnamefont {Blessing}},
  \bibinfo {author} {\bibfnamefont {J.}~\bibnamefont {Devlin}}, \bibinfo
  {author} {\bibfnamefont {J.}~\bibnamefont {Harrington}}, \bibinfo {author}
  {\bibfnamefont {T.}~\bibnamefont {Higuchi}}, \bibinfo {author} {\bibfnamefont
  {J.}~\bibnamefont {Morgner}}, \bibinfo {author} {\bibfnamefont
  {C.}~\bibnamefont {Smorra}}, \bibinfo {author} {\bibfnamefont
  {E.}~\bibnamefont {Wursten}}, \bibinfo {author} {\bibfnamefont
  {M.}~\bibnamefont {Bohman}}, \bibinfo {author} {\bibfnamefont
  {M.}~\bibnamefont {Wiesinger}},  \emph {et~al.},\ }\href@noop {} {\bibfield
  {journal} {\bibinfo  {journal} {Physical review letters}\ }\textbf {\bibinfo
  {volume} {122}},\ \bibinfo {pages} {043201} (\bibinfo {year}
  {2019})}\BibitemShut {NoStop}%
\bibitem [{\citenamefont {Mavadia}\ \emph {et~al.}(2014)\citenamefont
  {Mavadia}, \citenamefont {Stutter}, \citenamefont {Goodwin}, \citenamefont
  {Crick}, \citenamefont {Thompson},\ and\ \citenamefont
  {Segal}}]{mavadia2014optical}%
  \BibitemOpen
  \bibfield  {author} {\bibinfo {author} {\bibfnamefont {S.}~\bibnamefont
  {Mavadia}}, \bibinfo {author} {\bibfnamefont {G.}~\bibnamefont {Stutter}},
  \bibinfo {author} {\bibfnamefont {J.}~\bibnamefont {Goodwin}}, \bibinfo
  {author} {\bibfnamefont {D.}~\bibnamefont {Crick}}, \bibinfo {author}
  {\bibfnamefont {R.}~\bibnamefont {Thompson}}, \ and\ \bibinfo {author}
  {\bibfnamefont {D.}~\bibnamefont {Segal}},\ }\href@noop {} {\bibfield
  {journal} {\bibinfo  {journal} {Physical Review A}\ }\textbf {\bibinfo
  {volume} {89}},\ \bibinfo {pages} {032502} (\bibinfo {year}
  {2014})}\BibitemShut {NoStop}%
\bibitem [{\citenamefont {Schneider}(2017)}]{schneider2017300}%
  \BibitemOpen
  \bibfield  {author} {\bibinfo {author} {\bibfnamefont {G.}~\bibnamefont
  {Schneider}},\ }\emph {\bibinfo {title} {300 ppt Measurement of the Proton
  g-Factor}},\ \href@noop {} {Ph.D. thesis},\ \bibinfo  {school} {Johannes
  Gutenberg-Universit{\"a}t Mainz} (\bibinfo {year} {2017})\BibitemShut
  {NoStop}%
\bibitem [{\citenamefont {Mooser}\ \emph {et~al.}(2018)\citenamefont {Mooser},
  \citenamefont {Rischka}, \citenamefont {Schneider}, \citenamefont {Blaum},
  \citenamefont {Ulmer},\ and\ \citenamefont {Walz}}]{mooser2018new}%
  \BibitemOpen
  \bibfield  {author} {\bibinfo {author} {\bibfnamefont {A.}~\bibnamefont
  {Mooser}}, \bibinfo {author} {\bibfnamefont {A.}~\bibnamefont {Rischka}},
  \bibinfo {author} {\bibfnamefont {A.}~\bibnamefont {Schneider}}, \bibinfo
  {author} {\bibfnamefont {K.}~\bibnamefont {Blaum}}, \bibinfo {author}
  {\bibfnamefont {S.}~\bibnamefont {Ulmer}}, \ and\ \bibinfo {author}
  {\bibfnamefont {J.}~\bibnamefont {Walz}},\ }in\ \href@noop {} {\emph
  {\bibinfo {booktitle} {Journal of Physics: Conference Series}}},\ Vol.\
  \bibinfo {volume} {1138}\ (\bibinfo {organization} {IOP Publishing},\
  \bibinfo {year} {2018})\ p.\ \bibinfo {pages} {012004}\BibitemShut {NoStop}%
\bibitem [{\citenamefont {Ulmer}\ \emph {et~al.}(2011)\citenamefont {Ulmer},
  \citenamefont {Rodegheri}, \citenamefont {Blaum}, \citenamefont {Kracke},
  \citenamefont {Mooser}, \citenamefont {Quint},\ and\ \citenamefont
  {Walz}}]{ulmer2011observation}%
  \BibitemOpen
  \bibfield  {author} {\bibinfo {author} {\bibfnamefont {S.}~\bibnamefont
  {Ulmer}}, \bibinfo {author} {\bibfnamefont {C.~C.}\ \bibnamefont
  {Rodegheri}}, \bibinfo {author} {\bibfnamefont {K.}~\bibnamefont {Blaum}},
  \bibinfo {author} {\bibfnamefont {H.}~\bibnamefont {Kracke}}, \bibinfo
  {author} {\bibfnamefont {A.}~\bibnamefont {Mooser}}, \bibinfo {author}
  {\bibfnamefont {W.}~\bibnamefont {Quint}}, \ and\ \bibinfo {author}
  {\bibfnamefont {J.}~\bibnamefont {Walz}},\ }\href@noop {} {\bibfield
  {journal} {\bibinfo  {journal} {Physical review letters}\ }\textbf {\bibinfo
  {volume} {106}},\ \bibinfo {pages} {253001} (\bibinfo {year}
  {2011})}\BibitemShut {NoStop}%
\bibitem [{\citenamefont {Brownnutt}\ \emph {et~al.}(2015)\citenamefont
  {Brownnutt}, \citenamefont {Kumph}, \citenamefont {Rabl},\ and\ \citenamefont
  {Blatt}}]{brownnutt2015ion}%
  \BibitemOpen
  \bibfield  {author} {\bibinfo {author} {\bibfnamefont {M.}~\bibnamefont
  {Brownnutt}}, \bibinfo {author} {\bibfnamefont {M.}~\bibnamefont {Kumph}},
  \bibinfo {author} {\bibfnamefont {P.}~\bibnamefont {Rabl}}, \ and\ \bibinfo
  {author} {\bibfnamefont {R.}~\bibnamefont {Blatt}},\ }\href@noop {}
  {\bibfield  {journal} {\bibinfo  {journal} {Reviews of modern Physics}\
  }\textbf {\bibinfo {volume} {87}},\ \bibinfo {pages} {1419} (\bibinfo {year}
  {2015})}\BibitemShut {NoStop}%
\bibitem [{\citenamefont {Turchette}\ \emph {et~al.}(2000)\citenamefont
  {Turchette}, \citenamefont {King}, \citenamefont {Leibfried}, \citenamefont
  {Meekhof}, \citenamefont {Myatt}, \citenamefont {Rowe}, \citenamefont
  {Sackett}, \citenamefont {Wood}, \citenamefont {Itano}, \citenamefont
  {Monroe} \emph {et~al.}}]{turchette2000heating}%
  \BibitemOpen
  \bibfield  {author} {\bibinfo {author} {\bibfnamefont {Q.~A.}\ \bibnamefont
  {Turchette}}, \bibinfo {author} {\bibfnamefont {B.}~\bibnamefont {King}},
  \bibinfo {author} {\bibfnamefont {D.}~\bibnamefont {Leibfried}}, \bibinfo
  {author} {\bibfnamefont {D.}~\bibnamefont {Meekhof}}, \bibinfo {author}
  {\bibfnamefont {C.}~\bibnamefont {Myatt}}, \bibinfo {author} {\bibfnamefont
  {M.}~\bibnamefont {Rowe}}, \bibinfo {author} {\bibfnamefont {C.}~\bibnamefont
  {Sackett}}, \bibinfo {author} {\bibfnamefont {C.}~\bibnamefont {Wood}},
  \bibinfo {author} {\bibfnamefont {W.}~\bibnamefont {Itano}}, \bibinfo
  {author} {\bibfnamefont {C.}~\bibnamefont {Monroe}},  \emph {et~al.},\
  }\href@noop {} {\bibfield  {journal} {\bibinfo  {journal} {Physical Review
  A}\ }\textbf {\bibinfo {volume} {61}},\ \bibinfo {pages} {063418} (\bibinfo
  {year} {2000})}\BibitemShut {NoStop}%
\bibitem [{\citenamefont {Daniilidis}\ \emph {et~al.}(2011)\citenamefont
  {Daniilidis}, \citenamefont {Narayanan}, \citenamefont {M{\"o}ller},
  \citenamefont {Clark}, \citenamefont {Lee}, \citenamefont {Leek},
  \citenamefont {Wallraff}, \citenamefont {Schulz}, \citenamefont
  {Schmidt-Kaler},\ and\ \citenamefont
  {H{\"a}ffner}}]{daniilidis2011fabrication}%
  \BibitemOpen
  \bibfield  {author} {\bibinfo {author} {\bibfnamefont {N.}~\bibnamefont
  {Daniilidis}}, \bibinfo {author} {\bibfnamefont {S.}~\bibnamefont
  {Narayanan}}, \bibinfo {author} {\bibfnamefont {S.~A.}\ \bibnamefont
  {M{\"o}ller}}, \bibinfo {author} {\bibfnamefont {R.}~\bibnamefont {Clark}},
  \bibinfo {author} {\bibfnamefont {T.~E.}\ \bibnamefont {Lee}}, \bibinfo
  {author} {\bibfnamefont {P.~J.}\ \bibnamefont {Leek}}, \bibinfo {author}
  {\bibfnamefont {A.}~\bibnamefont {Wallraff}}, \bibinfo {author}
  {\bibfnamefont {S.}~\bibnamefont {Schulz}}, \bibinfo {author} {\bibfnamefont
  {F.}~\bibnamefont {Schmidt-Kaler}}, \ and\ \bibinfo {author} {\bibfnamefont
  {H.}~\bibnamefont {H{\"a}ffner}},\ }\href@noop {} {\bibfield  {journal}
  {\bibinfo  {journal} {New Journal of Physics}\ }\textbf {\bibinfo {volume}
  {13}},\ \bibinfo {pages} {013032} (\bibinfo {year} {2011})}\BibitemShut
  {NoStop}%
\bibitem [{\citenamefont {Poulsen}\ \emph {et~al.}(2012)\citenamefont
  {Poulsen}, \citenamefont {Miroshnychenko},\ and\ \citenamefont
  {Drewsen}}]{poulsen2012efficient}%
  \BibitemOpen
  \bibfield  {author} {\bibinfo {author} {\bibfnamefont {G.}~\bibnamefont
  {Poulsen}}, \bibinfo {author} {\bibfnamefont {Y.}~\bibnamefont
  {Miroshnychenko}}, \ and\ \bibinfo {author} {\bibfnamefont {M.}~\bibnamefont
  {Drewsen}},\ }\href@noop {} {\bibfield  {journal} {\bibinfo  {journal}
  {Physical Review A}\ }\textbf {\bibinfo {volume} {86}},\ \bibinfo {pages}
  {051402} (\bibinfo {year} {2012})}\BibitemShut {NoStop}%
\bibitem [{\citenamefont {Berlin}\ \emph {et~al.}()\citenamefont {Berlin},
  \citenamefont {Liu}, \citenamefont {Pospelov},\ and\ \citenamefont
  {Ramani}}]{blprfuture}%
  \BibitemOpen
  \bibfield  {author} {\bibinfo {author} {\bibfnamefont {A.}~\bibnamefont
  {Berlin}}, \bibinfo {author} {\bibfnamefont {H.}~\bibnamefont {Liu}},
  \bibinfo {author} {\bibfnamefont {M.}~\bibnamefont {Pospelov}}, \ and\
  \bibinfo {author} {\bibfnamefont {H.}~\bibnamefont {Ramani}},\ }\href@noop {}
  {\bibinfo  {journal} {Upcoming}\ }\BibitemShut {NoStop}%
\bibitem [{\citenamefont {Neufeld}\ \emph {et~al.}(2018)\citenamefont
  {Neufeld}, \citenamefont {Farrar},\ and\ \citenamefont
  {McKee}}]{Neufeld:2018slx}%
  \BibitemOpen
\bibfield  {journal} {  }\bibfield  {author} {\bibinfo {author} {\bibfnamefont
  {D.~A.}\ \bibnamefont {Neufeld}}, \bibinfo {author} {\bibfnamefont {G.~R.}\
  \bibnamefont {Farrar}}, \ and\ \bibinfo {author} {\bibfnamefont {C.~F.}\
  \bibnamefont {McKee}},\ }\href {\doibase 10.3847/1538-4357/aad6a4} {\bibfield
   {journal} {\bibinfo  {journal} {Astrophys. J.}\ }\textbf {\bibinfo {volume}
  {866}},\ \bibinfo {pages} {111} (\bibinfo {year} {2018})},\ \Eprint
  {http://arxiv.org/abs/1805.08794} {arXiv:1805.08794 [astro-ph.CO]}
  \BibitemShut {NoStop}%
\bibitem [{\citenamefont {Warring}\ \emph {et~al.}(2013)\citenamefont
  {Warring}, \citenamefont {Ospelkaus}, \citenamefont {Colombe}, \citenamefont
  {Brown}, \citenamefont {Amini}, \citenamefont {Carsjens}, \citenamefont
  {Leibfried},\ and\ \citenamefont {Wineland}}]{warring2013techniques}%
  \BibitemOpen
  \bibfield  {author} {\bibinfo {author} {\bibfnamefont {U.}~\bibnamefont
  {Warring}}, \bibinfo {author} {\bibfnamefont {C.}~\bibnamefont {Ospelkaus}},
  \bibinfo {author} {\bibfnamefont {Y.}~\bibnamefont {Colombe}}, \bibinfo
  {author} {\bibfnamefont {K.~R.}\ \bibnamefont {Brown}}, \bibinfo {author}
  {\bibfnamefont {J.}~\bibnamefont {Amini}}, \bibinfo {author} {\bibfnamefont
  {M.}~\bibnamefont {Carsjens}}, \bibinfo {author} {\bibfnamefont
  {D.}~\bibnamefont {Leibfried}}, \ and\ \bibinfo {author} {\bibfnamefont
  {D.~J.}\ \bibnamefont {Wineland}},\ }\href@noop {} {\bibfield  {journal}
  {\bibinfo  {journal} {Physical Review A}\ }\textbf {\bibinfo {volume} {87}},\
  \bibinfo {pages} {013437} (\bibinfo {year} {2013})}\BibitemShut {NoStop}%
\bibitem [{\citenamefont {Micke}\ \emph {et~al.}(2020)\citenamefont {Micke},
  \citenamefont {Leopold}, \citenamefont {King}, \citenamefont {Benkler},
  \citenamefont {Spie{\ss}}, \citenamefont {Schmoeger}, \citenamefont
  {Schwarz}, \citenamefont {L{\'o}pez-Urrutia},\ and\ \citenamefont
  {Schmidt}}]{micke2020coherent}%
  \BibitemOpen
  \bibfield  {author} {\bibinfo {author} {\bibfnamefont {P.}~\bibnamefont
  {Micke}}, \bibinfo {author} {\bibfnamefont {T.}~\bibnamefont {Leopold}},
  \bibinfo {author} {\bibfnamefont {S.}~\bibnamefont {King}}, \bibinfo {author}
  {\bibfnamefont {E.}~\bibnamefont {Benkler}}, \bibinfo {author} {\bibfnamefont
  {L.}~\bibnamefont {Spie{\ss}}}, \bibinfo {author} {\bibfnamefont
  {L.}~\bibnamefont {Schmoeger}}, \bibinfo {author} {\bibfnamefont
  {M.}~\bibnamefont {Schwarz}}, \bibinfo {author} {\bibfnamefont {J.~C.}\
  \bibnamefont {L{\'o}pez-Urrutia}}, \ and\ \bibinfo {author} {\bibfnamefont
  {P.}~\bibnamefont {Schmidt}},\ }\href@noop {} {\bibfield  {journal} {\bibinfo
   {journal} {Nature}\ }\textbf {\bibinfo {volume} {578}},\ \bibinfo {pages}
  {60} (\bibinfo {year} {2020})}\BibitemShut {NoStop}%
\bibitem [{\citenamefont {Cao}\ \emph {et~al.}(2016)\citenamefont {Cao},
  \citenamefont {Zhang}, \citenamefont {Shang}, \citenamefont {Cui},
  \citenamefont {Yuan}, \citenamefont {Chao}, \citenamefont {Wang},
  \citenamefont {Shu},\ and\ \citenamefont {Huang}}]{cao2016transportable}%
  \BibitemOpen
  \bibfield  {author} {\bibinfo {author} {\bibfnamefont {J.}~\bibnamefont
  {Cao}}, \bibinfo {author} {\bibfnamefont {P.}~\bibnamefont {Zhang}}, \bibinfo
  {author} {\bibfnamefont {J.}~\bibnamefont {Shang}}, \bibinfo {author}
  {\bibfnamefont {K.}~\bibnamefont {Cui}}, \bibinfo {author} {\bibfnamefont
  {J.}~\bibnamefont {Yuan}}, \bibinfo {author} {\bibfnamefont {S.}~\bibnamefont
  {Chao}}, \bibinfo {author} {\bibfnamefont {S.}~\bibnamefont {Wang}}, \bibinfo
  {author} {\bibfnamefont {H.}~\bibnamefont {Shu}}, \ and\ \bibinfo {author}
  {\bibfnamefont {X.}~\bibnamefont {Huang}},\ }\href@noop {} {\bibfield
  {journal} {\bibinfo  {journal} {arXiv preprint arXiv:1607.03731}\ } (\bibinfo
  {year} {2016})}\BibitemShut {NoStop}%
\bibitem [{\citenamefont {Delehaye}\ and\ \citenamefont
  {Lacro{\^u}te}(2018)}]{delehaye2018single}%
  \BibitemOpen
  \bibfield  {author} {\bibinfo {author} {\bibfnamefont {M.}~\bibnamefont
  {Delehaye}}\ and\ \bibinfo {author} {\bibfnamefont {C.}~\bibnamefont
  {Lacro{\^u}te}},\ }\href@noop {} {\bibfield  {journal} {\bibinfo  {journal}
  {Journal of Modern Optics}\ }\textbf {\bibinfo {volume} {65}},\ \bibinfo
  {pages} {622} (\bibinfo {year} {2018})}\BibitemShut {NoStop}%
\bibitem [{\citenamefont {Gellesch}\ \emph {et~al.}(2020)\citenamefont
  {Gellesch}, \citenamefont {Jones}, \citenamefont {Barron}, \citenamefont
  {Singh}, \citenamefont {Sun}, \citenamefont {Bongs},\ and\ \citenamefont
  {Singh}}]{gellesch2020transportable}%
  \BibitemOpen
  \bibfield  {author} {\bibinfo {author} {\bibfnamefont {M.}~\bibnamefont
  {Gellesch}}, \bibinfo {author} {\bibfnamefont {J.}~\bibnamefont {Jones}},
  \bibinfo {author} {\bibfnamefont {R.}~\bibnamefont {Barron}}, \bibinfo
  {author} {\bibfnamefont {A.}~\bibnamefont {Singh}}, \bibinfo {author}
  {\bibfnamefont {Q.}~\bibnamefont {Sun}}, \bibinfo {author} {\bibfnamefont
  {K.}~\bibnamefont {Bongs}}, \ and\ \bibinfo {author} {\bibfnamefont
  {Y.}~\bibnamefont {Singh}},\ }\href@noop {} {\bibfield  {journal} {\bibinfo
  {journal} {Advanced Optical Technologies}\ }\textbf {\bibinfo {volume} {9}},\
  \bibinfo {pages} {313} (\bibinfo {year} {2020})}\BibitemShut {NoStop}%
\bibitem [{\citenamefont {Carney}\ \emph {et~al.}(2021)\citenamefont {Carney},
  \citenamefont {H\"affner}, \citenamefont {Moore},\ and\ \citenamefont
  {Taylor}}]{Carney:2021irt}%
  \BibitemOpen
  \bibfield  {author} {\bibinfo {author} {\bibfnamefont {D.}~\bibnamefont
  {Carney}}, \bibinfo {author} {\bibfnamefont {H.}~\bibnamefont {H\"affner}},
  \bibinfo {author} {\bibfnamefont {D.~C.}\ \bibnamefont {Moore}}, \ and\
  \bibinfo {author} {\bibfnamefont {J.~M.}\ \bibnamefont {Taylor}},\ }\href
  {\doibase 10.1103/PhysRevLett.127.061804} {\bibfield  {journal} {\bibinfo
  {journal} {Phys. Rev. Lett.}\ }\textbf {\bibinfo {volume} {127}},\ \bibinfo
  {pages} {061804} (\bibinfo {year} {2021})},\ \Eprint
  {http://arxiv.org/abs/2104.05737} {arXiv:2104.05737 [quant-ph]} \BibitemShut
  {NoStop}%
\bibitem [{\citenamefont {Lang}\ and\ \citenamefont
  {Kohn}(1971)}]{lang1971theory}%
  \BibitemOpen
  \bibfield  {author} {\bibinfo {author} {\bibfnamefont {N.}~\bibnamefont
  {Lang}}\ and\ \bibinfo {author} {\bibfnamefont {W.}~\bibnamefont {Kohn}},\
  }\href@noop {} {\bibfield  {journal} {\bibinfo  {journal} {Physical Review
  B}\ }\textbf {\bibinfo {volume} {3}},\ \bibinfo {pages} {1215} (\bibinfo
  {year} {1971})}\BibitemShut {NoStop}%
\end{thebibliography}%


%
\end{document}